\documentclass[11pt,twoside,nohyper]{pallis}
\usepackage{epsfig}
\usepackage{latexsym}
\usepackage{euscript}
\usepackage{amssymb}
\usepackage{pstricks}
\usepackage{multirow,Bigstrut}
\usepackage{fancyhed}

\newcommand{\ftn}{\footnotesize}
\newcommand{\nsz}{\normalsize}
\newcommand{\ssz}{\scriptsize}

\newcommand{\tr}{{\mbox{\sf\ssz T}}}
\newcommand{\Tr}{\mbox{\sf Tr}}
\newcommand{\ntt}{\mbox{$\nu_\tau$}}

\def\beq{\begin{equation}}
\def\eeq{\end{equation}}
\def\bea{\begin{eqnarray}}
\def\eea{\end{eqnarray}}

\def\to{\rightarrow}

\def\llgm{\left\lgroup\matrix}
\def\rrgm{\right\rgroup}

\newcommand\sigv[2]{\langle v\sigma_{#1#2}\rangle}

\newcommand{\sff}{\mbox{$\tilde{f}$}}
\newcommand{\chai}{\mbox{${\tilde \chi_i^\pm}$}}

\newcommand{\nti}{\mbox{${\tilde\chi_i^0}$}}

\newcommand{\bdh}{\mbox{
$\llgm{h^+_2 & h^0_1 \cr h^0_2 & h^-_1 }\rrgm$}}

\newcommand{\bdhp}{\mbox{
$\llgm{h^{\prime +}_2 & h^{\prime 0}_1\cr h^{\prime 0}_2 &
h^{\prime -}_1}\rrgm$}}

\newcommand{\bdhpb}{\mbox{
$\llgm{\bar h^{\prime +}_2 &\bar h^{\prime 0}_1\cr\bar h^{\prime
0}_2 &\bar h^{\prime -}_1}\rrgm$}}

\newcommand{\Fi}{\mbox{
$\matrix{\llgm{d_r&-u_r}\rrgm\cr\llgm{e_r&-\nu_r}\rrgm}$}}

\newcommand{\Fqc}{\mbox{
$\llgm{u^c_r\cr d^c_r}\rrgm$}}

\newcommand{\Flc}{\mbox{
$\llgm{\nu^c_r\cr e^c_r}\rrgm$}}

\newcommand{\Hbc}{\mbox{
$\matrix{\llgm{\bar u^c_H&\bar d^c_H}\rrgm\cr\llgm{\bar
\nu^c_H&\bar e^c_H}\rrgm}$}}

\newcommand{\Hqc}{\mbox{
$\llgm{u^c_H\cr d^c_H}\rrgm$}}

\newcommand{\Hlc}{\mbox{
$\llgm{\nu^c_H\cr e^c_H}\rrgm$}}

\newcommand{\pf}{\mbox{
$\llgm{\phi&0\cr 0&-\phi}\rrgm$ }}

\newcommand{\bpf}{\mbox{
$\llgm{\bar\phi & 0\cr 0 & -\bar\phi}\rrgm$ }}
\preprint{\tt SISSA-19/2003/EP}

\title{\huge Y{\Large UKAWA} Q{\Large UASI}-U{\Large NIFICATION\\
AND} N{\Large EUTRALINO} R{\Large ELIC} D\Large ENSITY\\ \sl
\large Invited talk at the EUnet ``Supersymmetry and the Early
Universe''\\ mid-term meeting, Oxford, 26-29 September 2002}

\author{\Large C. P\nsz ALLIS \\
SISSA/ISAS, Via Beirut 2-4, 34013 Trieste, ITALY\\
\email{pallis@sissa.it}}
\author{\Large M.E. G\nsz\'OMEZ\\
Departamento de F\'{\i}sica and \\ Grupo Te\'orico de F\'{\i}sica
de Part\'{\i}culas, Instituto Superior T\'ecnico,
\\ Av. Rovisco Pais, 1049-001 Lisboa, PORTUGAL \\
\email{mgomez@cfif.ist.utl.pt}}

\abstract{The construction of a Supersymmetric Grand Unified Model
based on the Pati-Salam gauge group is briefly reviewed and the
low energy consequences of the derived asymptotic Yukawa
quasi-unification conditions are examined. In the framework of the
resulting Constrained Minimal Supersymmetric Standard Model, the
cosmological relic density of the bino-like LSP is calculated and
the results are explicitly compared with {\tt micrOMEGAs}. In
addition to the Cold Dark Matter constraint, restrictions on the
parameter space, arising from the Higgs boson masses, the SUSY
corrections to $b$-quark mass, the muon anomalous magnetic moment
and the inclusive decay $b\rightarrow s\gamma$ are, also,
investigated. For $\mu>0$, a wide and natural range of parameters
is allowed. On the contrary, the $\mu<0$ case not only is
disfavored from the present experimental data on the muon
anomalous magnetic moment, but also, it can be excluded from the
combination of the Cold Dark Matter and BR($b\to s \gamma$)
requirements.}

\begin{document}

\setcounter{page}{1}
\pagestyle{fancyplain}

\addtolength{\headheight}{.5cm}

\rhead[\fancyplain{}{ \bf \thepage}]{\fancyplain{}{ Y{\ftn UKAWA}
Q{\ftn UASI}-U{\ftn NIFICATION AND} N{\ftn EUTRALINO} R{\ftn ELIC}
D{\ftn ENSITY}}} \lhead[\fancyplain{}{
\leftmark}]{\fancyplain{}{\bf \thepage}} \cfoot{}

\section{I{\ftn NTRODUCTION}}\label{intro}

\hspace{.562cm} This review is based on Refs. \cite{qcdm}. We
briefly describe a Supersymmetric (SUSY) Grand Unified Theory
(GUT) which predicts a set of asymptotic Yukawa quasi-unification
conditions (YQUCs) in sec. \ref{quasi} and then we show how we can
restrict the parameter space of the resulting Constrained Minimal
Supersymmetric Standard Model (CMSSM) using a number of
Cosmo-Phenomenological requirements (sec. \ref{cdms}-\ref{pheno}).
Particular emphasis will be given in the Neutralino Relic Density
calculation in sec. \ref{cdms}. Some conclusions and open issues
will close this presentation in sec. \ref{cncl}.

\section{Y{\ftn UKAWA} Q{\ftn UASI}-U{\ftn NIFICATION}}\label{quasi}

\hspace{.562cm} The motivation of our construction is given in
subsec. \ref{yus} and a brief description of the model building
follows in subsec. \ref{model}. Finally, in subsec. \ref{qcmssm},
the mass parameters of the resulting CMSSM versions are derived.

\subsection{T{\ssz HE} Y{\ssz UKAWA} U{\ssz NIFICATION}
H{\ssz YPOTHESIS}}\label{yus}

\hspace{.562cm} We focus on a SUSY GUT based on the Pati Salam
(PS) gauge group $(SU(4)_{\rm c}\otimes SU(2)_L$ $\otimes
SU(2)_R={\bf 4}_{\rm c}{\bf 2}_L{\bf 2}_R)$ described in detail in
Ref. \cite{jean}. The relevant field content of the Model is
presented in the Table 1, where the components, the
representations, the transformations and the extra global charges
of the various superfields are, also, shown. The left handed
quarks and leptons superfields of every generation $r$ are
accommodated in a single pair of superfields $F_r, F^c_r$. The PS
symmetry can be spontaneously broken down to Standard Model (SM)
one through the vacuum expectation values (vevs) which the
superfields $H^c, \bar H^c$ acquire in the direction of the right
handed neutrinos, $\nu^c_H, \bar\nu^c_H$.

In the simplest realization of this model, as it is proposed by
Antoniadis and Leontaris in Ref. \cite{leontaris}, the electroweak
doublets $H_1, H_2$ are exclusively contained in the bidoublet
superfield $h$, which can be written:
\beq h=\llgm{h_2&h_1}\rrgm,\>\>\mbox{with}\>\>h_{1}=\llgm{h^0_1
\cr h^-_1}\rrgm\>\>\mbox{and}\>\> h_{2}=\llgm{h^+_2 \cr
h^0_2}\rrgm\cdot\eeq
With these assumptions, the model predicts Yukawa unification (YU)
at GUT scale, $M_{\rm GUT}$ ($M_{\rm GUT}$ is determined by the
requirement of gauge coupling unification):
\beq h_t(M_{\rm GUT})=h_b(M_{\rm GUT}) =h_\tau(M_{\rm GUT})
=y_{33}, \label{exact} \eeq
since the Yukawa coupling terms of the resulting version of MSSM
originate from a unique term of the underlying GUT, as follows:
\beq y_{33}\>F_3\>h\>F_3^c\>\>\ni\>\> y_{33}\>(H_2^\tr i \tau_2
Q\; t^c+H_1^\tr  i \tau_2 Q\; b^c + H_1^\tr i\tau_2 L\;
\tau^c),~~\mbox{since}~~H_{1[2]}=h_{1[2]}. \label{ffy}  \eeq
Here $Q, L \ni -i\tau_2 F_3^\tr $ are the well known $SU(2)_L$
doublets quarks and leptons of the third family and the self
explanatory substitutions $(t,b,\tau)$ in the position of
$(u_3,d_3,e_3)$ appeared in the Table 1 are made. Also, $\tau_2$
is the second Pauli matrix, $\ni$ means ``belongs to'' and the
brackets are used by applying disjunctive correspondence.

However, applying this scheme in the context of the CMSSM
\cite{Cmssm} and given the top and tau experimental masses (which
naturally restrict the $\tan\beta\simeq50$), the pre-diction of
the (asymptotic) YU leads to a contra-diction with the
experimental bounds on $b$-quark mass. This is due to the fact
that in the large $\tan\beta$ regime, the tree level $b$-quark
mass, $m_b$, receives sizeable (about 20$\%$) SUSY corrections
\cite{copw} $\Delta m_b(M_{\rm SUSY})$, evaluated at a SUSY
breaking scale $M_{\rm SUSY}$ (see sec. \ref{qcmssm}). These
corrections arise from sbottom-gluino (mainly) and stop-chargino
loops \cite{copw, pierce, susy} and have the sign of $\mu$ (with
the standard sign convention of Ref. \cite{sugra}). Their size is
a 1-loop {\it exact} quantity, as it is explained in Ref.
\cite{susy}. Hence, for $\mu>[<]~0$, the SUSY corrections drive
the corrected (which is essentially the so called running
\cite{susy}) $b$-quark mass, $m^{\rm c}_b$, at a low scale $M_Z$,
\beq m^{\rm c}_b(M_Z)=m_b(M_Z)\Big(1+\Delta m_b(M_{\rm SUSY})\Big)
\eeq
a lot above [a little below] than its $95\%$ confidence level
(c.l.) experimental range. This is derived by appropriately
\cite{qcdm} evolving the corresponding range for the pole
$b$-quark mass, $m_b(m_b)$ up to $M_Z$ scale with
$\alpha_s(M_Z)\simeq 0.1185$, in accord with the analysis in Ref.
\cite{baermb}:
\beq m_b(m_b) = 4.25\pm 0.3\>{\rm
GeV}\>\>\>\stackrel{}{\Longrightarrow}\>\>\>
 m^{\rm c}_b(M_Z) = 2.88\pm 0.2\>{\rm GeV} \label{mbrg}\eeq
Concluding, for both signs of $\mu$ the YU assumption leads to an
unacceptable $b$-quark mass. The discrepancy is less sizable for
$\mu<0$, so we can say that YU ``prefers'' \cite{nath} $\mu<0$.

The usual strategy to accommodate the latter discrepancy is the
introduction of several kinds of nonuniversalities in the scalar
\cite{king, raby, baery} and/or gaugino \cite{nath} sector of
MSSM, without an exact conservation of the YU. On the contrary, in
Ref. \cite{qcdm}, this problem is addressed in the context of the
PS GUT, without need of invoking the departure from the CMSSM 
\cite{Cmssm} universality. The Higgs sector of the model 
is extended so, that the electroweak Higgs in Eq. (\ref{ffy})
receive subdominant contributions from other 
representations, too. As a consequence, a
moderate violation of the YU is obtained, which can allow an
acceptable $b$-quark mass, even with universal boundary
conditions.

\addtolength{\textheight}{1.cm}
\newpage

\begin{table}[!ht]
\caption{\bf R\ssz ELEVANT \nsz F\ssz IELD \nsz C\ssz ONTENT \nsz}
\begin{center}
\begin{tabular}{|ccccccc|}
\hline \multirow{2}{0.55in}{\bf F\ssz IELDS}&\multirow{2}{1.03in}
{\bf C\ssz OMPONENTS}&{\bf R\ssz EPRESE-}&\multirow{2}{0.73in}{\bf
T\ssz RASFOR-}&\multicolumn{3}{c|}{ {\bf G\ssz LOBAL} }
\\
\multicolumn{2}{|c}{}&{\bf\ssz NTATIONS}
&\multirow{2}{0.65in}{\bf\ssz MATIONS} &\multicolumn{3}{c|}{ {\bf
S\ssz YMMETRIES}}
\\
\multicolumn{2}{|c}{(Color Index Suppressed)}&{${\bf 4}_{\rm
c}{\bf 2}_L{\bf 2}_R$} &&{$R$} &{$ PQ$} &{$Z^{mp}_2$}
\\\hline \hline
\multicolumn{2}{|c}{\bf M{\ssz ATTER}  F{\ssz IELDS}} &
\multicolumn{5}{c|}{}
\\ \hline \multicolumn{7}{|c|}{} \\
{$F_r$} &{$\Fi$}&{$({\bf 4, 2, 1})$}&$F_rU_L^{\dagger}U^\tr_{\rm
c}$ & $1/2$ & $-1$ &$1$
\\
\multicolumn{7}{|c|}{}
 \\
{$F^c_r$} &{$\Fqc,\Flc$} & {$({\bf \bar 4, 1, 2})$}&$U_{\rm
c}^\ast U_R^\ast F^c_r$ &{ $1/2$ }&{$0$}&{$-1$}
\\ \multicolumn{7}{|c|}{}
 \\
\multicolumn{2}{|c}{  ($r=1,2,3$: Generation Index)} &
\multicolumn{5}{c|}{}\\ \hline
\multicolumn{2}{|c}{\bf H\ssz IGGS  \nsz F\ssz IELDS} &
\multicolumn{5}{c|}{}
\\ \hline
\multicolumn{7}{|c|}{}
 \\
{$H^c$} &{$\Hqc,\Hlc$} &{$({\bf \bar 4, 1, 2})$}& $U_{\rm c}^\ast
U_R^\ast H^c$ &{$0$}&{$0$} & { $0$}
\\
\multicolumn{7}{|c|}{}
 \\
{$\bar H^c$}&{$\Hbc$}&$({\bf 4, 1, 2})$& $\bar{H}^cU^\tr_R
U^\tr_{\rm c}$&{$0$}&{$0$}&{$0$}
 \\
\multicolumn{7}{|c|}{}
 \\
{$h$} &{$\bdh$} & {$({\bf 1, 2, 2})$}&$U_LhU^\tr_R$ &$0$ &$1$ &$0$
\\
\multicolumn{7}{|c|}{}
 \\ \hline
\multicolumn{2}{|c}{\bf E\ssz XTRA \nsz H\ssz IGGS \nsz F\ssz
IELDS } & \multicolumn{5}{c|}{}
\\ \hline
\multicolumn{7}{|c|}{}
\\
$h^{\prime}$&{$\bdhp$} &{$({\bf 15, 2, 2})$} & $U_{\rm c}^\ast
U_Lh^{\prime}U^\tr_RU^\tr_{\rm c}$ & $0$ & $1$ &$0$
\\
\multicolumn{7}{|c|}{}
 \\
 $\bar h^{\prime}$&{$\bdhpb$}&{$({\bf 15, 2, 2})$} &
$U_{\rm c}U_L\bar{h}^{\prime}U^\tr_RU_{\rm c}^\dagger$ & $1$ &
$-1$ &$0$
\\
\multicolumn{7}{|c|}{}
\\
$\phi$&$\pf^{(\ast)}$ &$({\bf 15, 1, 3})$& $U_{\rm c}U_R\phi
U_R^\dagger U_{\rm c}^\dagger$ & $0$ & $0$ &$0$
\\
\multicolumn{7}{|c|}{}
\\
$\bar\phi$&$\bpf^{(\ast)}$&{$({\bf 15, 1, 3})$} &$U_{\rm c}U_R\bar
\phi U_R^\dagger U_{\rm c}^\dagger$ & $1$ & $0$ &$0$
\\ \cline{4-4}
\multicolumn{3}{|c}{}&$U_{\rm c}\in SU(4)_{\rm
c}$&\multicolumn{3}{c|}{}
\\
&\multicolumn{1}{c}{$^{(\ast)}$SM neutral} &&$U_{L}\in
SU(2)_{L}$&\multicolumn{3}{c|}{}\\
%\multicolumn{7}{|c|}{}
%\\
&\multicolumn{1}{c}{content} & &$U_{R}\in SU(2)_{R}$
&\multicolumn{3}{c|}{}\\
\hline
\end{tabular}
\end{center}
\end{table}

\addtolength{\textheight}{-1.cm}
\newpage

\subsection{M{\ssz ODEL} C{\ssz ONSTRUCTION}}\label{model}

\hspace{.562cm} From the discussion of the previous section, we
can induce that a small deviation from the YU will be enough for
an appropriate prediction of $b$-quark mass when $\mu<0$, while
for $\mu>0$, a more pronounced deviation is needed. According to
the first paper in Ref. \cite{qcdm}, this can be achieved by
introducing some extra higgs fields, as follows:

Since the matter fields product has the following structure under
${\bf 4}_{\rm c}{\bf 2}_L{\bf 2}_R$:
\beq F_3F_3^c\ni{\bf (4, 2, 1)} \otimes {\bf (\bar 4, 1, 2)}={\bf
(1, 2, 2)}\oplus {\bf (15, 2, 2)}\label{fs}\eeq
and ${\bf 15}_{SU(4)}\ni {\bf 1}_{SU(3)}$, it is possible the
addition of two higgs superfields:
\beq h^\prime,\> \bar h^\prime\>\>\in\>\> {\bf (15, 2,
2)}.\label{hs}\eeq
The field $h^\prime$ can couple to $F_3F_3^c$ and $\bar h^\prime$
can give mass to the color non singlets through a mixing term
$m\bar h^\prime h^\prime$ with $m\sim M_{\rm
GUT}\simeq2\times10^{16}~{\rm GeV}$, in accord with the imposed
global symmetries (see Table 1). Other possible mixing term
suppressed by the string scale $M_S\simeq5\times10^{17}~{\rm
GeV}$, being non renormalizable, is:
\bea \bar H^c H^c \bar h^\prime h/M_S \>\>\ni\>\>\Big({\bf
15}\otimes {\bf 15},~{\bf 1}\otimes {\bf 1},~({\bf 1}\oplus {\bf
3})\otimes({\bf 1}\oplus{\bf 3})\Big), \label{nonren} \eea
with the latter structure under ${\bf 4}_{\rm c}{\bf 2}_L{\bf
2}_R$, since for the 2 participants of the product, we get:
\bea\bar H^c H^c~[\bar h^\prime h]~~\in~~\Big({\bf 15}\oplus {\bf
1}~[{\bf 15}],~{\bf1}~[{\bf1}\oplus{\bf3}],~{\bf1}\oplus{\bf3}
\Big). \nonumber \eea
In the term of Eq. (\ref{nonren}), there are 2 couplings as
regards the $SU(2)_R$: A singlet, $\lambda_{\bf1}(\bar H^c
H^c)_{\bf 1}$ $\bar h^\prime h$ (since
${\bf1}\otimes{\bf1}={\bf1}$), and a triplet, $\lambda_{\bf3}(\bar
H^c H^c)_{\bf3}\bar h^\prime h$ (since ${\bf3} \otimes
{\bf3}={\bf1}\oplus {\bf3}\oplus{\bf5}$). As it turns out, the
singlet coupling provides us with an adequate deviation from the
YU for $\mu<0$. The necessary deviation for $\mu>0$ can be
obtained by a further enlargement of the Higgs sector, such that
contributions arising from renormalizable terms are allowed. By
introducing two additional Higgs fields:
\beq \phi,\> \bar \phi\>\> \in \>\>{\bf (15, 1, 3)}\label{phi}\eeq
an unsuppressed coupling $\lambda^\prime_{\bf 3} \phi \bar
h^\prime h$ (again since ${\bf3} \otimes {\bf3}={\bf1}\oplus
{\bf3}\oplus {\bf5}$) can be constructed. This overshadows the
coupling $\lambda_{\bf3}(\bar H^c H^c)_{\bf3}\bar h^\prime h$.
$\bar\phi$ is introduced to give superheavy masses to the color
non singlets in $\phi$ through a term $\bar\phi\phi$. Summarizing,
the important mixing terms are:
\bea \nonumber m\bar h^\prime h^\prime,\>\lambda_{{\bf 1}[{\bf
3}]} (\bar H^c H^c)_{{\bf 1}[{\bf3}]} \bar h^\prime
h/M_S~~\mbox{and}~~\lambda^\prime_{\bf 3} \phi \bar h^\prime h~~
\left(\lambda_{{\bf1}[{\bf3}]}, \lambda^\prime_{\bf 3}:~
\mbox{dimensionless constants}\right). \eea

Using the transformation properties of the several fields
indicated in the Table 1, and taking into account the unitarity of
the $SU(2)_{L,R}$ groups, we conclude that the previous mixing
terms correspond to the following invariant quantities under
${\bf4}_{\rm c}{\bf2}_L{\bf2}_R$, respectively:
\beq m\Tr\left(\bar h^{\prime}
i\tau_2h^{\prime\tr}i\tau_2\right),\>\>\frac{\lambda_{{\bf 1}[{\bf
3}]}}{M_S}\Tr\left(\bar h^{\prime} i\tau_2(\bar H^{c\tr}
H^{c\tr})_{{\bf1}[{\bf3}]} h^\tr
i\tau_2\right)\>\>\mbox{and}\>\>\lambda^\prime_{\bf
3}\Tr\left(\bar h^{\prime} i\tau_2\phi h^\tr i\tau_2 \right),
\label{expmix} \eeq
where the traces are taken with respect to the $SU(4)_{\rm c}$ and
$SU(2)_L$ indices.

At GUT scale, during the spontaneous breaking of PS gauge
symmetry, the fields $\bar H^c, H^c$ and $\phi$ acquire vevs in
the SM singlet direction. Therefore,
\beq \left(\langle\bar H^{c\tr}\rangle\langle
H^{c\tr}\rangle\right)_{{\bf1}[{\bf3}]}=v^2_{H^c}\left(T_{H^c},1,\frac{\sigma_{0[3]}}{2}
\right)\>\>\mbox{and}\>\>\langle\phi\rangle=v_\phi
\left(T^{15},1,\frac{\sigma_3}{\sqrt{2}} \right),\label{vevs}\eeq
where $v_{H^c,\phi}\sim M_{\rm GUT}$ and the three-fold structure
of these vevs as regards the ${\bf4}_{\rm c}{\bf2}_L{\bf2}_R$
group is ordered in the parenthesis by commas, with
\beq T_{H^c}={\sf diag}\Big(0,0,0,1\Big),~~T^{15}=
\frac{1}{2\sqrt{3}}\>{\sf
diag}\Big(1,1,1,-3\Big)~~\mbox{and}~~\sigma_{0[3]}={\sf
diag}\Big(1,[-]1\Big).\eeq
Expanding the superfields in Eq. (\ref{hs}) as linear combination
of the 15 generators $T^a$ of $SU(4)_{\rm c}$ with the
normalization $\Tr(T^aT^b)=\delta^{ab}$ and denoting the SM
singlet components with the superfield symbol, the following
identities can be easily proved:
\bea && \Tr\left(\bar h^{\prime} i\tau_2h^{\prime\tr}
i\tau_2\right)=\bar h^{\prime\tr}_1i\tau_2 h^\prime_2 +
h^{\prime\tr}_1i\tau_2\bar h^\prime_2+{\cal
O}(T^aT^a,~a<15),\label{idnta}
\\ && \Tr\left(\bar h^{\prime}
i\tau_2\left(T^{15}_{[H^c]},1,\sigma_{3[0]}\right)h^\tr
i\tau_2\right)=\left(\bar h^{\prime\tr}_1i\tau_2 h_2 -[+]
h^\tr_1i\tau_2\bar h^\prime_2\right)\left[-\sqrt{3}/2\right],
\label{idnt}\eea
where the notation of Eqs. (\ref{vevs}) was applied. Inserting
Eqs. (\ref{vevs}) in Eqs. (\ref{expmix}) and employing Eqs.
(\ref{idnta})-(\ref{idnt}), we get the mass terms:
\beq m\bar h^{\prime\tr}_1i\tau_2\left(h^{\prime}_2+
\alpha_2h_2\right)+m\left(h^{\prime \tr}_1+\alpha_1
h^\tr_1\right)i\tau_2\bar{h}^{\prime}_2, \label{superheavy}\eeq
where the the mixing effects are included in the following
coefficients:
\beq \alpha_{1[2]}=
\frac{1}{\sqrt{2}m}\left(-[+]\,\left(\lambda^\prime_{\bf
3}v_\phi-\frac{\sqrt{6}}{4M_S}\lambda_{\bf
3}v^2_{H^c}\right)-\frac{\sqrt{6}}{4M_S}\lambda_{\bf
1}v^2_{H^c}\right)\cdot \label{alphas} \eeq
It is obvious from Eq. (\ref{superheavy}) that we obtain 2
combinations of superheavy massive fields:
\bea\bar{h}^{\prime}_1,~H^{\prime}_1~~\mbox{and}~~
\bar{h}^{\prime}_2,~H^{\prime}_2,~~\mbox{where}~~
H^{\prime}_{i}=\frac{h^{\prime}_{i}+\alpha_{i}h_{i}}
{\sqrt{1+|\alpha_{i}|^2}},~i=1,2. \label{shs}\eea
The electroweak doublets, remaining massless at GUT scale, can be
interpreted as ``orthogonal'' to $H^{\prime}_{i}$ direction:
\beq H_i=\frac{-\alpha_i^*h^{\prime}_i+h_i}
{\sqrt{1+|\alpha_i|^2}} \label{elws}\eeq
\par
Solving Eqs. (\ref{elws}) and (\ref{shs}) as respect $h_i$ and
$h^\prime_i$, we obtain:
\beq h_i=\frac{H_i+\alpha^*_iH^{\prime}_i}{\sqrt{1+|\alpha_i|^2}}
~~\mbox{and}~~h^\prime_i=\frac{-\alpha_iH_i+H^{\prime}_i}
{\sqrt{1+|\alpha_i|^2}}\cdot~~~\eeq
Consequently, the Yukawa couplings terms, in contrast to Eq.
(\ref{ffy}), take the form:
$$ y_{33}\>F_3\>h\>F_3^c+2y^\prime_{33}\>F_3\>h^\prime\>F_3^c
\>\>\ni\>\>y_{33}\>F_3 \llgm{h_2+2\rho T^{15}h^\prime_2& h_1+2\rho
T^{15}h^\prime_1}\rrgm F_3^c \>\>\ni$$
$$y_{33}\>\left(
\frac{1-\rho\alpha_2/\sqrt{3}}{\sqrt{1+|\alpha_2|^2}}\>H_2^\tr
i\tau_2Q\>t^c
+\frac{1-\rho\alpha_1/\sqrt{3}}{\sqrt{1+|\alpha_1|^2}}\>H_1^\tr
i\tau_2Q\>b^c
+\frac{1+\sqrt{3}\rho\alpha_1}{\sqrt{1+|\alpha_1|^2}}\>H_1^\tr
i\tau_2L\>\tau^c\right),$$
where $\rho=y^\prime_{33}/y_{33}$ with $0<\rho<1$ and only color
singlets components of $h^\prime_i$ are shown. The fields
$H^\prime_i$ being superheavy mass states, contribute neither to
the running of renormalization group equations (RGEs) nor to the
masses of fermions. Consequently, the asymptotic exact YU in Eq.
(\ref{exact}) can be replaced by a set of asymptotic YQUCs:
\bea h_t(M_{\rm GUT}):h_b(M_{\rm GUT}):h_\tau(M_{\rm GUT})=
\left|\frac{1-\rho\alpha_2/\sqrt{3}}
{\sqrt{1+|\alpha_2|^2}}\right|:
\left|\frac{1-\rho\alpha_1/\sqrt{3}}
{\sqrt{1+|\alpha_1|^2}}\right|:
\left|\frac{1+\sqrt{3}\rho\alpha_1}
{\sqrt{1+|\alpha_1|^2}}\right|\cdot~~~~\label{qg}\eea
The deviation from the YU can be estimated, by defining the
following relative splittings:
\begin{equation}
\delta h_{b[\tau]}=\left(h_{b[\tau]}(M_{\rm GUT})-h_t(M_{\rm
GUT})\right)/h_t(M_{\rm GUT}). \label{splitting}
\end{equation}

%%%%%%%%%%%%%%%%%%%%%%%%%%%%%%%%%%%%%%%%%%%%%%%%%%%%%%%%%%%%%%%%%%%%
\begin{figure}[!ht]
\hspace*{-.71in}
\begin{minipage}{8in}
\epsfig{file=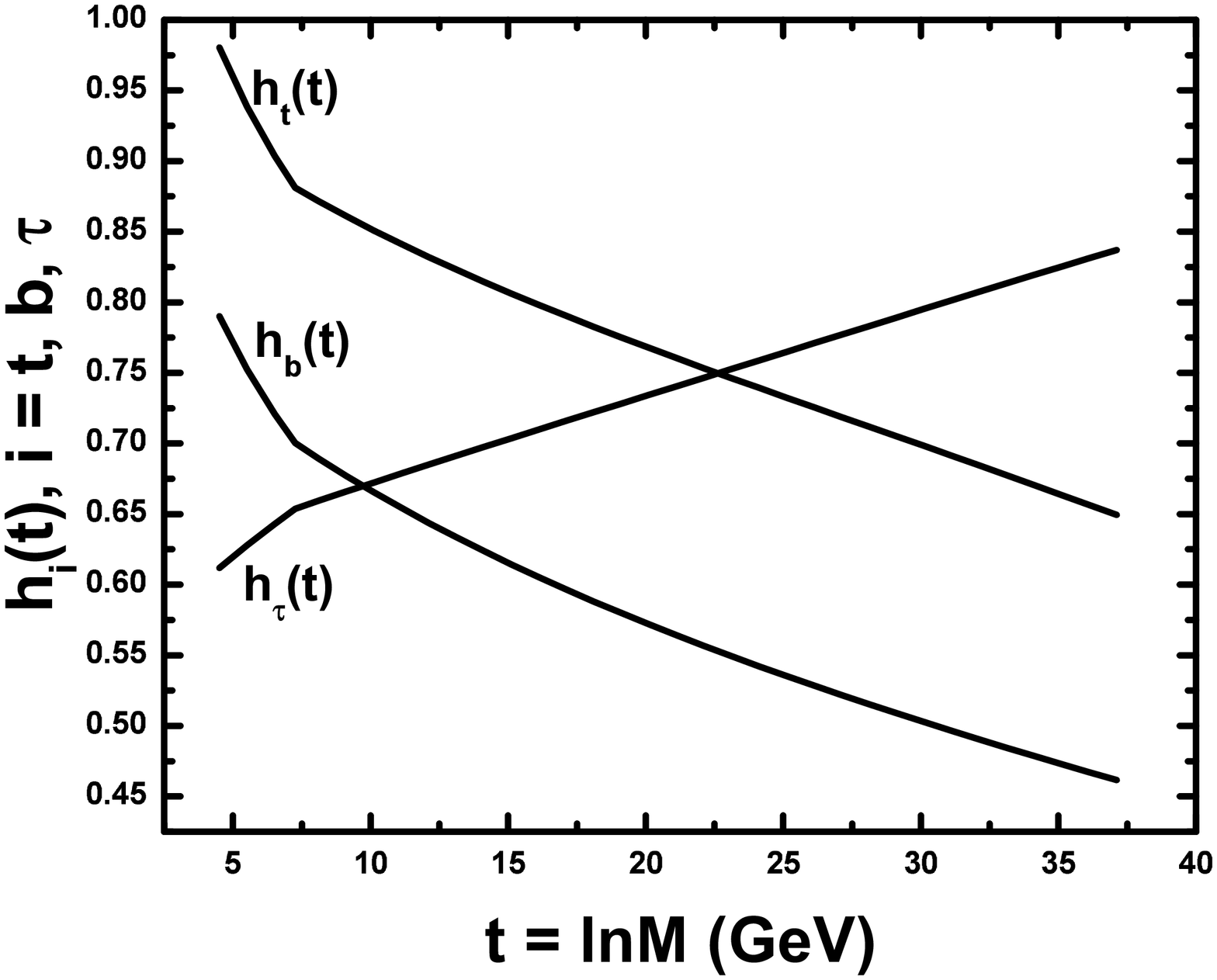,height=3.8in,angle=-90}
\hspace*{-1.37 cm}
\epsfig{file=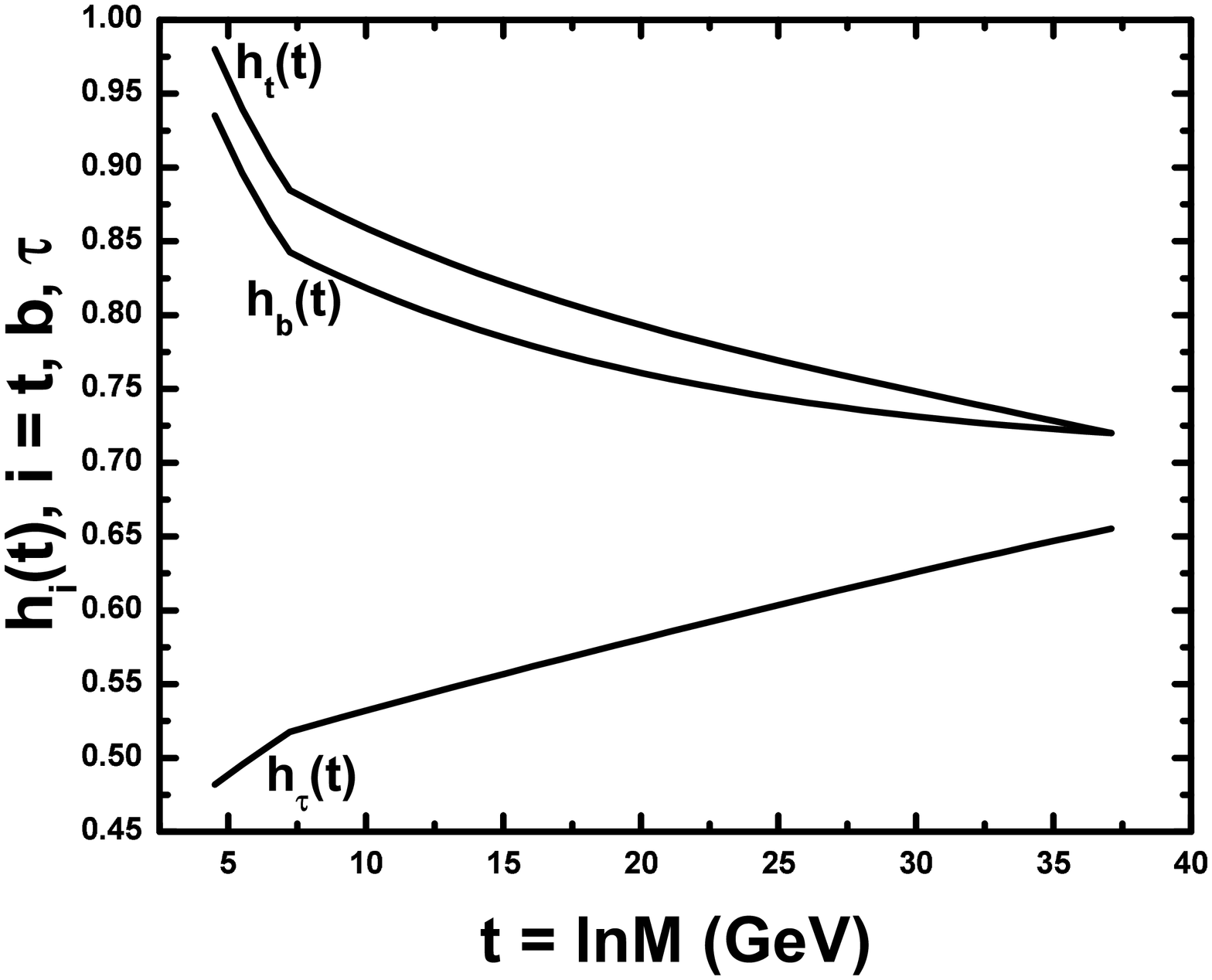,height=3.8in,angle=-90} \hfill
\end{minipage}
\hfill \caption{\sl A Yukawa couplings running from $M_{\rm GUT}$
to $M_Z$ for $m^{\rm c}_b(M_Z)=2.68~{\rm GeV}$, $\mu>[<]~0$,
$m_{\rm LSP}=297~[372]~{\rm GeV}$ and
$\Delta_{\tilde\tau_2}=1.5~[0]$ corresponding to $\tan\beta\simeq
59~[50]$ (left [right] plot).}\label{yuk}
\end{figure}
%%%%%%%%%%%%%%%%%%%%%%%%%%%%%%%%%%%%%%%%%

\subsection{CMSSM  W{\ssz ITH} Y{\ssz UKAWA} Q{\ssz UASI}-U{\ssz NIFICATION}}
\label{qcmssm}

\hspace{.562cm}  Below the $M_{\rm GUT}$, our particle content
reduces to this of MSSM. We assume universal soft SUSY breaking
scalar masses, $m_0$, gaugino masses, $M_{1/2}$ and trilinear
scalar couplings, $A_0$ at $M_{\rm GUT}$. Therefore, the resulting
MSSM is the so called CMSSM \cite{Cmssm} supplemented by a
suitable YQUC from the set in Eq. (\ref{qg}). With these initial
conditions, we run the MSSM RGEs \cite{cdm} between $M_{\rm GUT}$
and a common SUSY threshold $M_{\rm SUSY}\simeq(m_{\tilde
t_1}m_{\tilde t_2})^{1/2}$ ($\tilde t_{1,2}$ are the stop mass
eigenstates) determined in consistency with the SUSY spectrum. At
$M_{\rm SUSY}$ we impose radiative electroweak symmetry breaking,
evaluate the SUSY spectrum and incorporate the SUSY corrections to
$b$ and $\tau$ masses \cite{pierce, susy, king}. The latter
(almost 4$\%$) lead \cite{cdm} to a small de[in]-crease of
$\tan\beta$ for $\mu>[<]~0$.  From $M_{\rm SUSY}$ to $M_Z$, the
running of gauge and Yukawa couplings is continued using the SM
RGEs.

For presentation purposes, $M_{1/2}$ and $m_0$ can be replaced
\cite{cdm} by the lightest SUSY particle (LSP) mass, $m_{\rm LSP}$
and the relative mass splitting between it and the lightest stau
$\tilde\tau_2$, $\Delta_{\tilde\tau_2}$. For simplicity, we
restrict this presentation to the $A_0=0$ case (for $A_0\neq0$ see
Refs. \cite{qcdm, mario}). Our parameter ``transmutation'' is
shown schematically, as follows:
$${\rm sign}\mu,\>M_{1/2},\ m_0,\
A_0\>\>\stackrel{A_0=0}{\Longrightarrow}\>\>{\rm sign}\mu,\>m_{\rm
LSP},\
\Delta_{\tilde\tau_2}\>\>\>\left(\Delta_{\tilde\tau_2}=(m_{\tilde\tau_2}
-m_{\rm LSP })/m_{\rm LSP}\right)$$

We restrict ourselves to monoparametric YQUCs derived from the
general Eq. (\ref{qg}) with real $\alpha_{1,2}$. It is convenient
to define the quantity:
$$c=\rho\alpha_1/\sqrt{3}~~\mbox{with}~~-1/3<c<1$$
For any given $m^{\rm c}_b(M_Z)$ in Eq. (\ref{mbrg}) with fixed
top, $m_t(m_t)=166~{\rm GeV}$ and tau, $m^{\rm c
}_\tau(M_Z)=1.746~{\rm GeV}$ masses, we can determine the
parameters $c$ and $\tan\beta$ at $M_{\rm SUSY}$ so, that the YQUC
in Eq. (\ref{qg}) is satisfied. However, for the plots in Figs. 1,
2 and 3 we fix $m^{\rm c}_b(M_Z)=2.68~{\rm GeV}$, since this value
turns out to overlap the allowed areas for $\mu>0$ or give the
less restrictive upper bound for $\mu<0$ on the $m_{\rm
LSP}-\Delta_{\tilde\tau_2}$ plane (Figs. 4 and 5).

%%%%%%%%%%%%%%%%%%%%%%%%%%%%%%%%%%%%%%%%%%%%%%%%%%%%%%%%%%%%%%%%%%%%
\begin{figure}[t]
\hspace*{-.71in}
\begin{minipage}{8in}
\epsfig{file=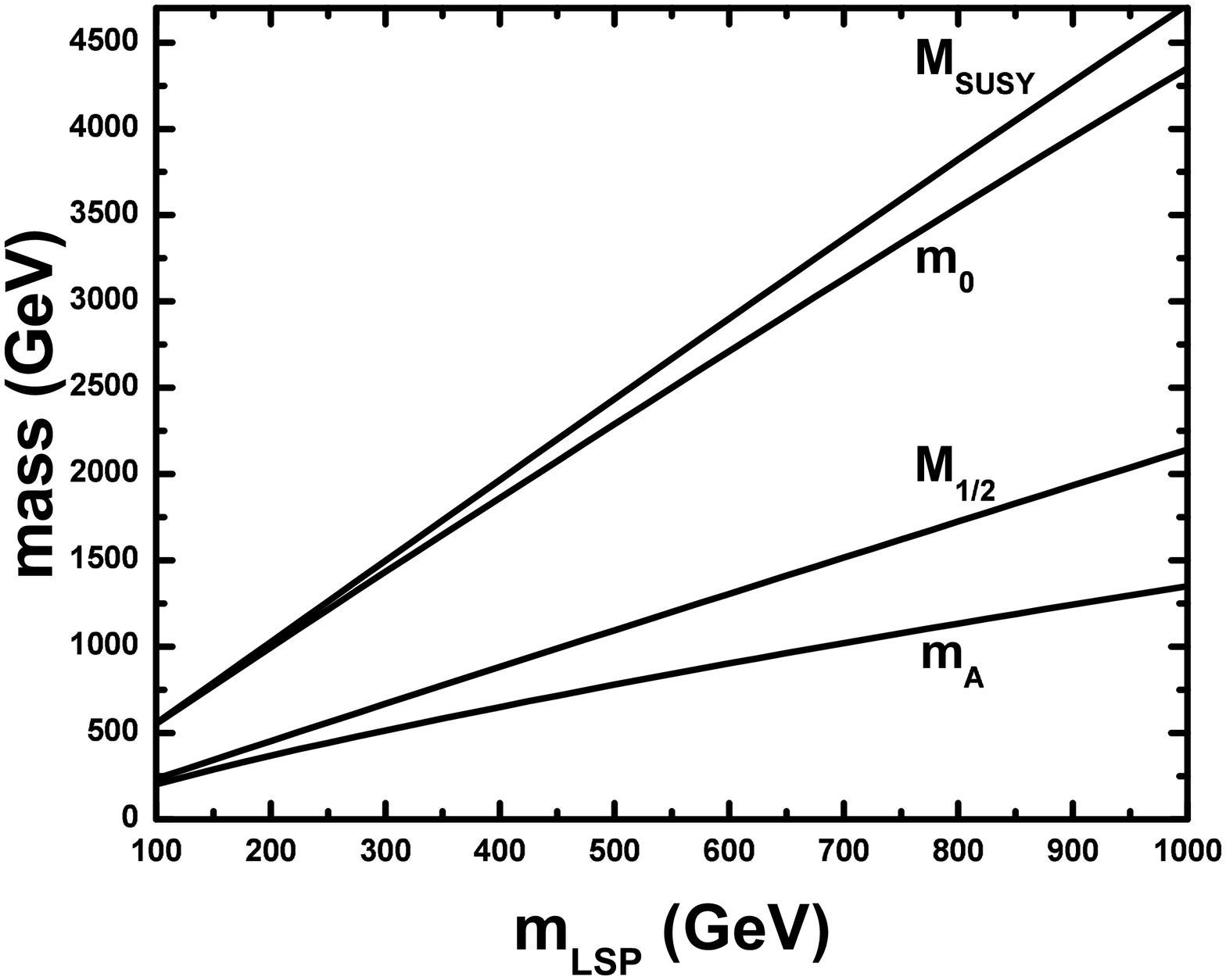,height=3.8in,angle=-90}
\hspace*{-1.37 cm}
\epsfig{file=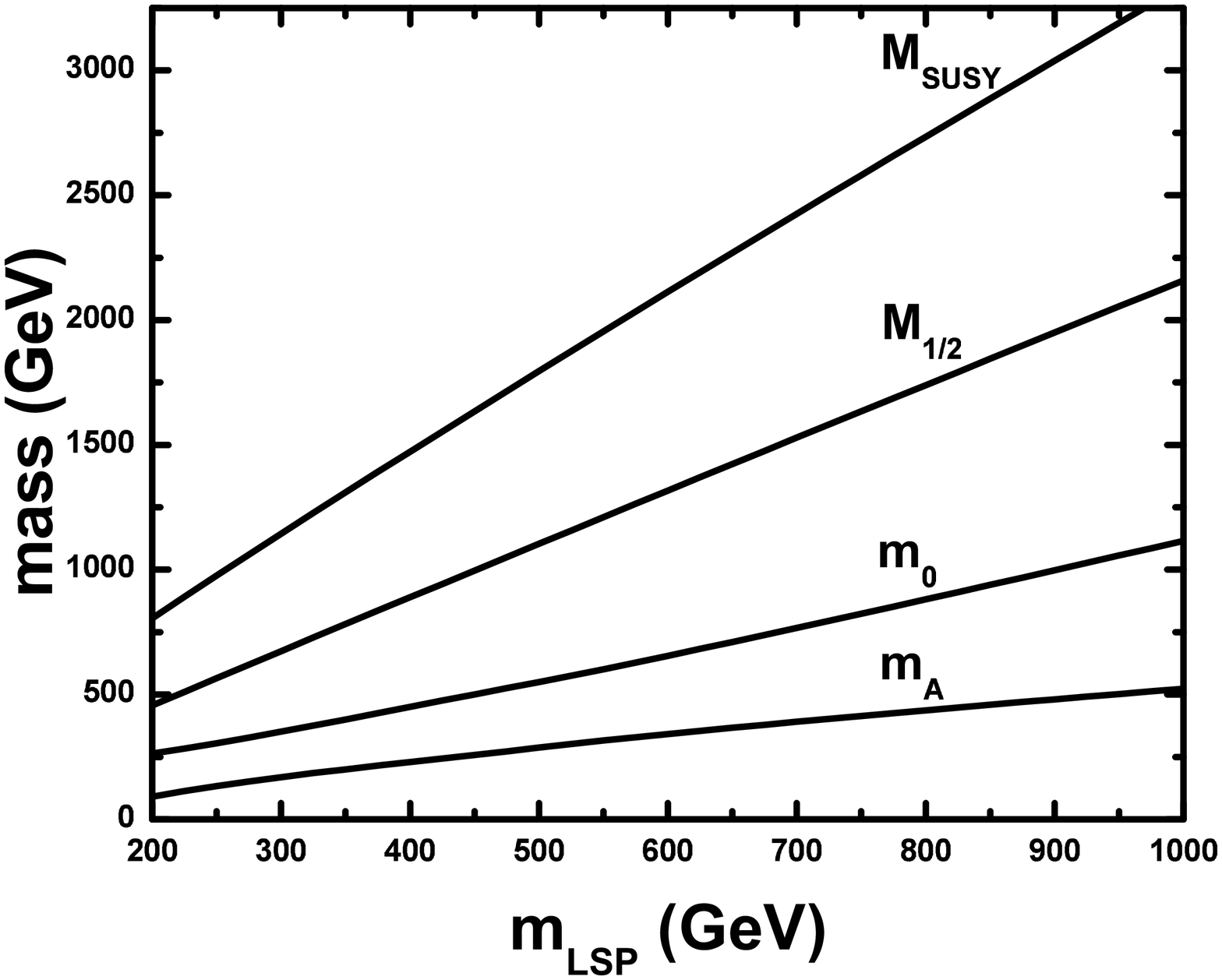,height=3.8in,angle=-90} \hfill
\end{minipage}
\hfill \caption[]{\sl The mass parameters $m_A$, $m_0$, $M_{1/2}$
and $M_{\rm SUSY}$ versus $m_{\rm LSP}$ for $m^{\rm
c}_b(M_Z)=2.68~{\rm GeV}$, $\mu>[<]~0$ and
$\Delta_{\tilde\tau_2}=1.5~[0]$ corresponding to $\tan\beta\simeq
59~[50]$ (left [right] plot).}\label{Mnx}
\end{figure}
%%%%%%%%%%%%%%%%%%%%%%%%%%%%%%%%%%%%%%%%%

The YQUC predicting  appropriate values for $m^{\rm c}_b(M_Z)$
relates the minimal Higgs content of the model described in
sec.~\ref{model} to the sign of $\mu$, as follows:

$\bullet$ For $\mu>0$ the fields of Eq.~(\ref{phi}) are needed.
Viable CMSSM spectrum is obtained for $\alpha_1=-\alpha
_2\>\sim\>1$ in Eq. (\ref{alphas}).  Therefore, Eq. (\ref{qg})
implies:
\beq h_t(M_{\rm GUT}):h_b(M_{\rm GUT}):h_\tau(M_{\rm
GUT})=|1+c|:|1-c|:|1+3c|.\eeq
In the left panel of Fig. \ref{yuk}, we present a typical Yukawa
couplings running from $M_{\rm GUT}$ to $M_Z$ for $m_{\rm
LSP}=297~{\rm GeV}$ and $\Delta_{\tilde\tau_2}=1.5$. In the left
panel of Fig. \ref{Mnx}, the values of $m_A$, $m_0$, $M_{1/2}$ and
$M_{\rm SUSY}$ versus $m_{\rm LSP}$ are, also, displayed for the
same $\Delta_{\tilde\tau_2}$. We observe that $M_{1/2}<m_0$ and
$2m_{\rm LSP}\simeq m_A$ for $m_{\rm LSP}<350~{\rm GeV}$. For
$2.68~{\rm GeV}\lesssim m^{\rm c}_b(M_Z)\lesssim 3.09~{\rm GeV}$,
the parameters $c$, $\delta h_{b[\tau]}$ and $\tan\beta$ range,
correspondingly:
\bea 0.19\gtrsim c\gtrsim0.13,\>\>0.32\gtrsim\delta
h_{\tau}=-\delta h_b=2c/(1+c)\gtrsim0.23\>\>\mbox{and}\>\>60
\gtrsim \tan\beta \gtrsim 57. \nonumber \eea

$\bullet$ For $\mu<0$, the inclusion of the fields in Eq.
(\ref{phi}) can be avoided. Interesting CMSSM spectrum is obtained
for $\alpha_1=\alpha _2\>\sim\>0.1$ in Eq. (\ref{alphas}).
Therefore, Eq. (\ref{qg}) implies:
\beq h_t(M_{\rm GUT}):h_b(M_{\rm GUT}):h_\tau(M_{\rm
GUT})=|1-c|:|1-c|:|1+3c|.\eeq
In the right panel of Fig. \ref{yuk}, we present a typical Yukawa
couplings running from $M_{\rm GUT}$ to $M_Z$ for $m_{\rm
LSP}=372~{\rm GeV}$ and $\Delta_{\tilde\tau_2}=0$. In the right
panel of Fig. \ref{Mnx}, the values of $m_A$, $m_0$, $M_{1/2}$ and
$M_{\rm SUSY}$ versus $m_{\rm LSP}$ are, also, displayed for the
same $\Delta_{\tilde\tau_2}$. We observe that $M_{1/2}\gg m_0$ and
$2m_{\rm LSP}\gg m_A$. For $2.68~{\rm GeV}\lesssim m^{\rm
c}_b(M_Z)\lesssim 3.09~{\rm GeV}$, the parameters $c$, $\delta
h_{b[\tau]}$ and $\tan\beta$ range, correspondingly:
\bea 0.01\lesssim -c\lesssim 0.06,\>\>0.04\lesssim-\delta
h_{\tau}=-4c/(1-c)\lesssim0.25,\>\>\delta
h_b=0\>\>\mbox{and}\>\>50\gtrsim\tan\beta\gtrsim47. \nonumber \eea

Worth mentioning is finally, that in our models (in contrast to
the ``traditional'' CMSSM version \cite{Cmssm}) $\tan\beta$ is not
a free parameter but a prediction of the applied YQUC.

\newpage
\section{N{\ftn EUTRALINO} R{\ftn ELIC} D{\ftn ENSITY}}
\label{cdms}

\hspace{.562cm} In the context of CMSSM, the LSP can be the
lightest neutralino. It naturally arises as a Cold Dark Matter
(CDM) candidate \cite{goldberg}. We require its relic density,
$\Omega_{\rm LSP}h^2$, not to exceed the upper bound derived from
DASI on the CDM abundance at $95\%$ c.l. \cite{dasi}:
\beq \Omega_{\rm CDM}h^2\lesssim0.22\label{cdmb}\eeq
For both signs of $\mu$ an upper bound on $m_{\rm LSP}$ can be
derived from this requirement. Some generic features of
$\Omega_{\rm LSP}h^2$ calculation in the CMSSM are exploited  in
subsec. \ref{cdmg}, and particular applications to the model under
consideration are exhibited in subsec. \ref{pole}, \ref{coan}.

\subsection{G{\ssz ENERAL} C{\ssz ONSIDERATIONS}} \label{cdmg}

\hspace{.562cm} In most of the CMSSM parameter space, the LSP is
almost a pure bino and $\Omega_{\rm LSP}h^2$ increases with
$m_{\rm LSP}$. Therefore, Eq. (\ref{cdmb}) sets an upper
approximate limit on its mass: $m_{\rm LSP}\lesssim200~{\rm GeV}$.
However, as it is pointed out in Refs. \cite{lah, ellis1}, a
substantial reduction of $\Omega_{\rm LSP}h^2$ can be achieved in
some regions of the parameter space, thanks to two reduction
``procedures'': The $A$-pole effect ($A$PE) and the coannihilation
mechanism (CAM). The first is activated for $\tan\beta>40~
[\simeq(30-35)]$ with $\mu>[<]~0$, where the presence of a
resonance ($2m_{\rm LSP}\simeq m_A$) in the $A$ Higgs mediated
LSP' s annihilation channel is possible. On the other hand, CAM is
applicable for any $\tan\beta$, both signs of $\mu$ but it
requires a mass proximity between LSP and the next-to-LSP, which
turns out to be the lightest stau, $\tilde\tau_2$, for
$\tan\beta>10$ \cite{cmssm, ellis2} and not too large values for
$A_0$ \cite{drees} or $m_0$ \cite{baer, darkn}.

Our model gives us the opportunity to discuss the operation of
both reduction ``procedures''. As it is induced from Fig. 2, for
$\mu>0$ there is a significant region with $A$PE, while for
$\mu<0$ the CAM is the only available reduction ``procedure''.
Furthermore, for $\mu<0$, the CAM is more strengthened, since due
to the larger $m_{\rm LSP}$, more coannihilation channels are
kinematically allowed than for $\mu>0$. For this reason some
technical details for each reduction ``procedure'' will be
presented separately in the two following subsections for $\mu>0$
and $\mu<0$ respectively.

We calculate $\Omega_{\rm LSP}h^2$, using {\tt micrOMEGAs}
\cite{micro}, which is one of the most complete publicly available
codes. This includes accurately thermally averaged exact
tree-level cross sections of all possible (co)annihilation
processes, and loop QCD corrections to the Higgs couplings into
fermions. The results of this code are checked in Refs. \cite{qcdm,
mario} with another public package {\sf DarkSUSY} \cite{dark} (not
the newest version \cite{darkn}) appropriately combined
\cite{qcdm} with the code used in Ref. \cite{cdm}. We found good
agreement when the Higgs couplings are treated at tree level. This
agreement persists, even with loop QCD corrections to these
couplings \cite{qcd}, provided an artificial $m_b(m_b)$ is used in
the defaults of {\sf DarkSUSY}, in order to mimic these
corrections.

In this talk, a new comparison is presented with an improved
version of the code (let name it, GLP) presented in Ref. \cite{cdm}.
A first, model independent improvement concerns the freeze out
procedure \cite{gelmini}. This is renewed, using variable values
for the number of relativistic degrees of freedom, $g_*$, whose a
precise estimation is obtained, employing the tables included in
{\tt micrOMEGAs} package. Details on other specific improvements
together with the comparisons will be displayed in the next
subsections.

\subsection{I{\ssz NCLUDING THE} {\sl A}-P{\ssz OLE} E{\ssz FFECT}}
\label{pole}

\begin{table}[!h]
\caption{\bf C\ssz OMPARISON OF \nsz {\boldmath $\Omega_{\rm
LSP}h^2$} C\ssz ALCULATIONS}
\begin{center}
\begin{tabular}{|l||c|c||c|c||c|c|} \hline
\multicolumn{7}{|c|}{\bf M\ssz ODEL \nsz P\ssz ARAMETERS}\\
\multicolumn{7}{|c|}{($\mu>0,~m^{\rm c}_b(M_Z)=2.68~{\rm
GeV},~A_0=0~{\rm GeV}$)}\\ \hline
$\tan\beta$&\multicolumn{2}{c||}{59}&\multicolumn{2}{|c||}{58.9}&
\multicolumn{2}{|c|}{58.8} \\
$M_{1/2}~({\rm GeV})$
&\multicolumn{2}{c||}{712}&\multicolumn{2}{c||}{663}&
\multicolumn{2}{c|}{616}  \\
$m_0~({\rm GeV})$
&\multicolumn{2}{c||}{1298}&\multicolumn{2}{c||}{1484}&
\multicolumn{2}{c|}{1638}  \\ \hline
$m_A~({\rm
GeV})$&\multicolumn{2}{c||}{560}&\multicolumn{2}{c||}{510}&
\multicolumn{2}{c|}{462} \\
$m_{\rm LSP}~({\rm GeV})$
&\multicolumn{2}{c||}{319}&\multicolumn{2}{c||}{297}&
\multicolumn{2}{c|}{276}\\
$\Delta_{\tilde\tau_2}$&\multicolumn{2}{c||}{1.0}&\multicolumn{2}{c||}{1.5}&
\multicolumn{2}{c|}{2.0} \\ \hline \hline
\multicolumn{7}{|c|}{\bf \nsz {\boldmath $\Omega_{\rm LSP}h^2$}
C\ssz ALCULATIONS}\\ \hline
\multirow{2}{1in}{\bf C\ssz ODE \nsz :}&{\tt
micr-}&\multirow{2}{0.31in}{GLP}&{\tt
micr-}&\multirow{2}{0.31in}{GLP}&{\tt
micr-}&\multirow{2}{0.31in}{GLP}
\\ &{\tt OMEGAs}& &{\tt OMEGAs}& &{\tt OMEGAs}&\\ \hline \hline
$\Gamma^{\rm tree}_A~({\rm GeV})$&66&64.4&59&58.4&54&52.6 \\
$\Gamma^{\rm QCD}_A~({\rm GeV})$&34&32.5&31&29.8&29&27.0\\
$\Gamma^{\rm SUSY}_A~({\rm GeV})$&-&24.9&-&23.3&-&21.5 \\ \hline
\hline
$\Omega_{\rm LSP}h^2|_{\rm
tree}$&0.124&0.117&0.123&0.116&0.122&0.115
\\
$\Omega_{\rm LSP}h^2|_{\rm QCD}$&0.221&0.218
&0.223&0.221&0.222&0.221
\\
$\Omega_{\rm LSP}h^2|_{\rm SUSY}$&-&0.270&-&0.271&-&0.267\\ \hline
\end{tabular}
\end{center}
\end{table}

In the absence of the CAM, $\Omega_{\rm LSP}h^2$ is inverse
proportional to the thermally averaged product of relative
velocity times the $\tilde\chi-\tilde\chi$ cross section,
$\sigv{\tilde\chi}{\tilde\chi}$. In the presence of the $A$PE,
this is enhanced, due to the $s$-channel $A$ exchange, $s(A)$, to
the down type fermion-antifermion pairs.

For a reliable calculation of $\Omega_{\rm LSP}h^2$ in this
regime, two points have to be taken into account: First, since the
low velocity expansion of $\sigv{\tilde\chi}{\tilde\chi}$ breaks
down \cite{gelmini} in the vicinity of poles, the full phase-space
integration for the $s(A[H])$ channels with fermions to final
states has to be performed, for masses in the interval
\cite{cmssm} $0.65<2m_{\tilde\chi}/m_{A[H]}<2$ (the $H$
contribution is p-wave suppressed \cite{cmssm, roberto}). Second,
a careful treatment of the relevant $A$-fermions couplings
$g_{Aff}$, and the corresponding decay widths is, also,
indispensable \cite{micro}. Since these are proportional to the
corresponding fermion mass, $m_f$, a rather accurate estimation of
their tree level \cite{qcd}, $g^{\rm tree}_{Aff}(M)$, loop QCD
corrected \cite{qcd}, $g^{\rm QCD}_{Aff}(M)$ and SUSY resummed
\cite{susy} QCD corrected $g^{\rm SUSY}_{Aff}(M)$ value at a scale
$M$ can be achieved, if $m_f$ is replaced by an effective fermion
mass. More explicitly,
\beq -\frac{g\>m_f}{2M_W}\tan\beta=\left\{\matrix{
%\begin{array}{rl}
g^{\rm tree}_{Aff}(M)\hfill ,  & m_f=m_f(m_f) \hfill \cr
g^{\rm QCD}_{Aff}(M)\hfill ,  & m_f=m_f(M)\Big(1+\Delta m_f(M_{\rm
SUSY})\Big)\Delta^{1/2}_{\rm QCD}(M) \hfill \cr
g^{\rm SUSY}_{Aff}(M)\hfill , & m_f=m_f(M)\Delta^{1/2}_{\rm
QCD}(M) \hfill \cr}
%\end{array}
\right. \eeq
with $\Delta_{\rm QCD}=1$ [given in Ref. \cite{qcd} (applying the
$m_A\gg m_f$ limit\footnote{We are grateful to Dr. P. Ullio for
bringing to our attention this point.})] for $f=\tau~[b]$
($g$ is the $SU(2)_L$ gauge coupling). Inserting these
effective couplings calculated at $m_A$ in the tree level formula
for the $A$-decay width \cite{qcd}, we reach the {\tt micrOMEGAs}
results in the case of tree level, $\Gamma^{\rm tree}_A$ and QCD
corrected, $\Gamma^{\rm QCD}_A$ $A$-decay width, within a 5$\%$
accuracy as it is shown in the Table 2. An estimation, also, can
be done for the SUSY improved value of the $A$-decay width,
$\Gamma^{\rm SUSY}_A$ (not included in the current version of {\tt
micrOMEGAs}).

For the inputs of Table 2, the relative contributions of the
annihilation processes to the $\Omega_{\rm LSP}h^2$ calculation
are as follows:
\bea \nonumber \tilde\chi\tilde\chi\rightarrow b\bar
b~[\tau\bar\tau]& ~~~~&93~[6]\%~~\mbox{at tree level,} \eea
with a 6$\%$ de[in]-crease at loop QCD level and a 8$\%$
de[in]-crease at SUSY improved loop QCD level. In the same Table,
we compare, also, the corresponding values for the tree level,
$\Omega_{\rm LSP}h^2|_{\rm tree}$ and loop QCD corrected,
$\Omega_{\rm LSP}h^2|_{\rm QCD}$, $\Omega_{\rm LSP}h^2$ with {\tt
micrOMEGAs}. An agreement within a (4-0)$\%$ accuracy is achieved.
Note that the various couplings in our code are calculated at
$2m_{\rm LSP}$ scale. Moreover, the 45$\%$ increase of
$\Omega_{\rm LSP}h^2$ because of
the loop QCD corrections to $g_{Aff}$'s and $A$-decay width
(firstly noticed in Ref. \cite{micro}) is impressively reproduced.
An estimation can be done, as well, for the result on $\Omega_{\rm
LSP}h^2$, if SUSY corrections are included, $\Omega_{\rm
LSP}h^2|_{\rm SUSY}$. An almost $18\%$ increase is expected. This
statement can not be reliably checked through {\tt micrOMEGAs},
since the user is able to introduce these corrections only to
$A$-decay width and not to $g_{Aff}$'s \cite{micro}, too.
%
%%%%%%%%%%%%%%%%%%%%%%%%%%%%%%%%%%%%%%%%%%%%%%%%%%%%%%%%%%%%%%%%%%%%
\begin{figure}[!t]
\hspace*{-.71in}
\begin{minipage}{8in}
\epsfig{file=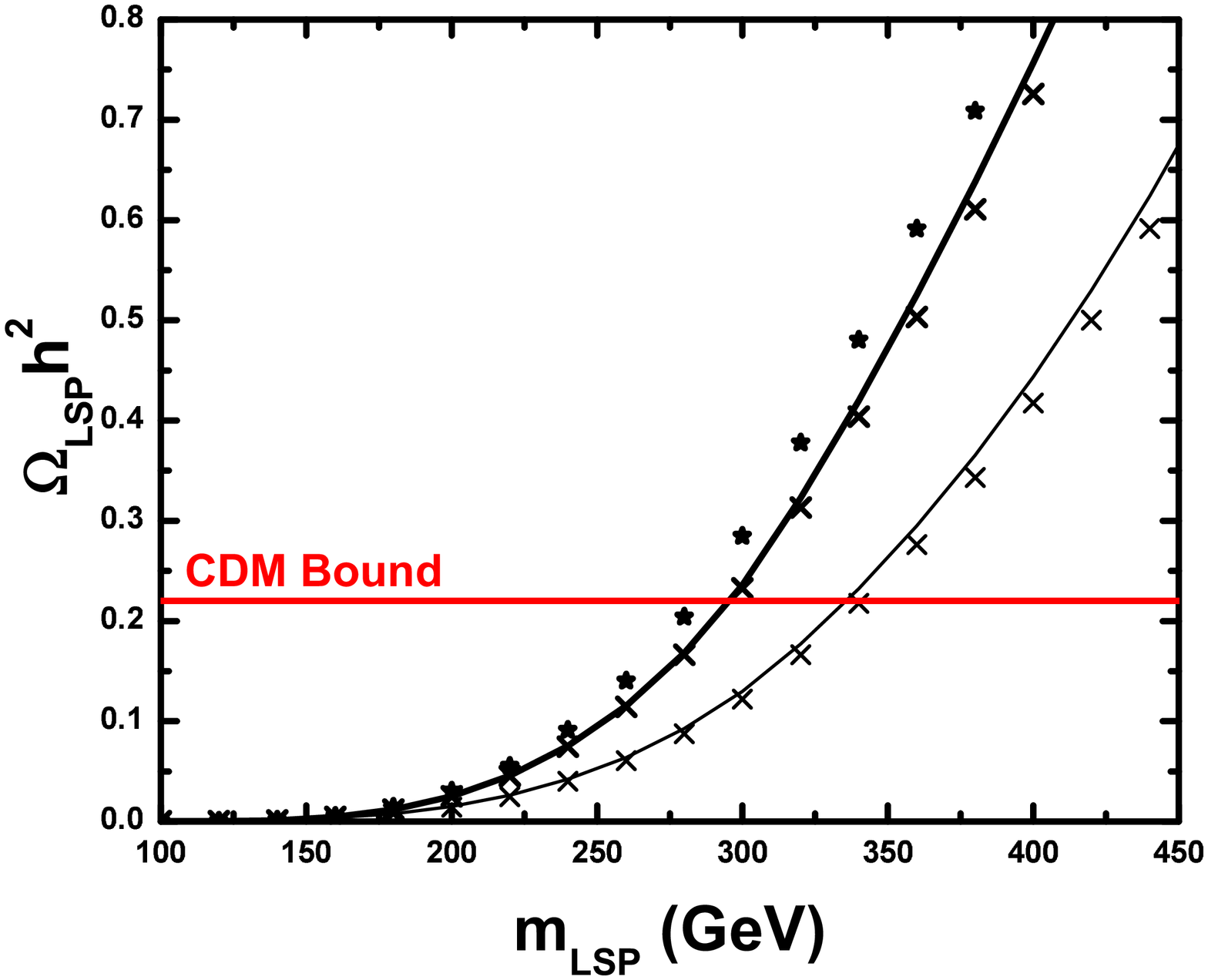,height=3.8in,angle=-90}
\hspace*{-1.37 cm}
\epsfig{file=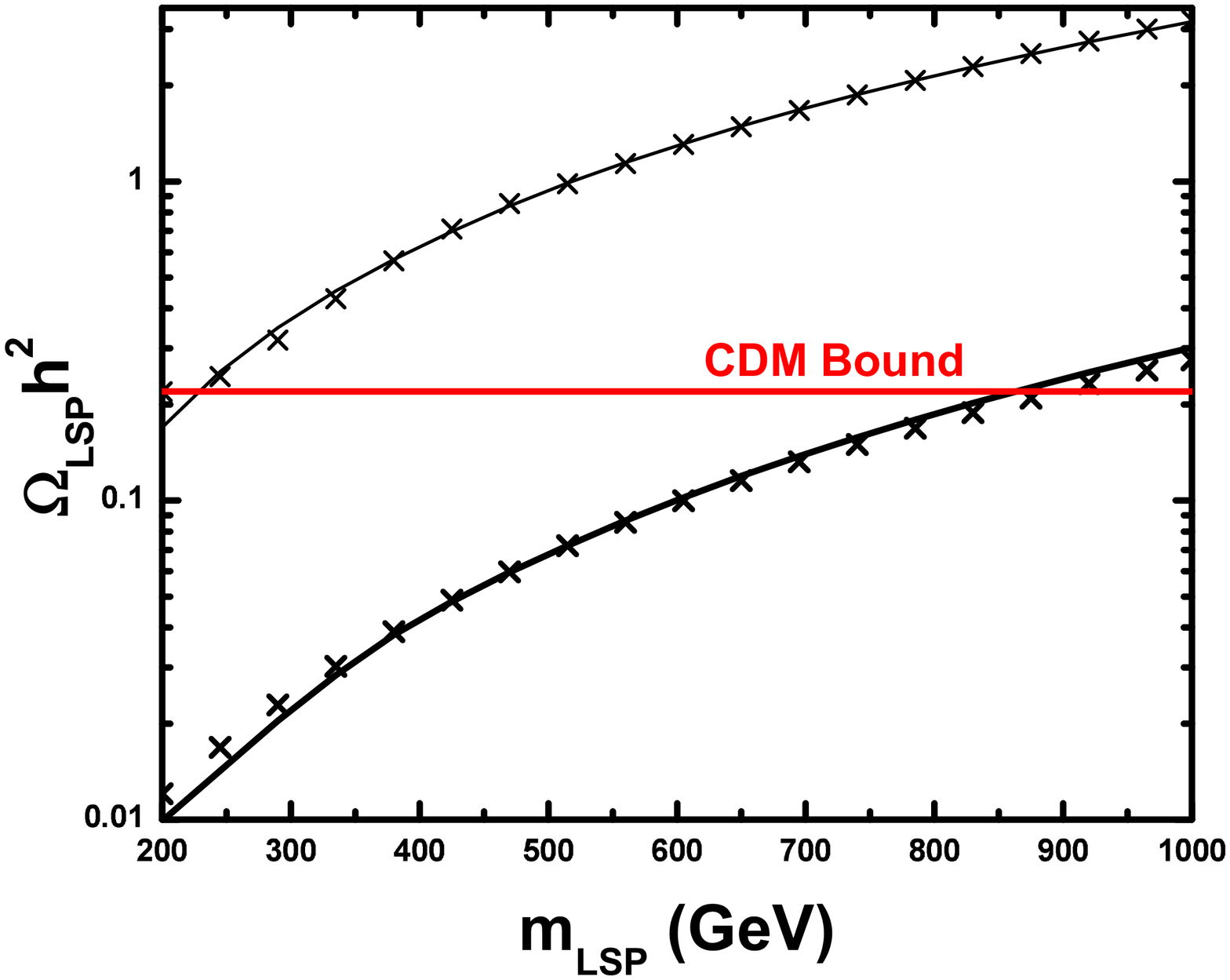,height=3.8in,angle=-90}
\hfill
\end{minipage}
\hfill \caption[]{\sl $\Omega_{\rm LSP}h^2$ versus $m_{\rm LSP}$
for $m^{\rm c}_b(M_Z)=2.68~{\rm GeV},\mu>[<]~0$ and
$\Delta_{\tilde\tau_2}=1.5~[0]$ (left [right] plot). The thick
(faint) solid line is obtained by {\tt micrOMEGAs} with (without)
loop QCD corrections [coannihilations] while the thick (faint)
crosses by our code. The stars are obtained by our code, including
SUSY corrections, too. The upper bound on $\Omega_{\rm
LSP}h^2~(=0.22)$ is also depicted.}\label{om}
\end{figure}
%%%%%%%%%%%%%%%%

A further comparison is displayed in Fig. 3 (left plot), where we
depict $\Omega_{\rm LSP}h^2$ versus $m_{\rm LSP}$ for
$\mu>0,\Delta_{\tilde\tau_2}=1.5$ and $m_b(M_Z)=2.68~{\rm GeV}$.
$\Omega_{\rm LSP}h^2$ is calculated by {\tt micrOMEGAs} [our code]
with (thick solid line [crosses]) or without (faint solid line
[crosses]) the inclusion of the loop QCD corrections to $g_{Aff}$
couplings. The stars are obtained by our code, including the SUSY
corrections, too.

%The dashed line is drawn using the data of our code including the
%SUSY corrections, too.

\newpage
%%%%%%%%%%%%%%%%%%%%%%%%%%%%%%%%%%%%%%%%%

\subsection{I{\ssz NCLUDING} B{\ssz INO}-S{\ssz TAU} C{\ssz OANNIHILATIONS}}
\label{coan}

\begin{table}[!h]
\caption{\bf I\ssz NCLUDED \nsz F\ssz EYNMAN \nsz D\ssz IAGRAMS}
\begin{center}
\begin{tabular}{|c|c|c|}
\hline {\bf I\ssz NITIAL \nsz S\ssz TATE }&{\bf F\ssz INAL \nsz
S\ssz TATE} &{\bf I\ssz NTERACTION \nsz C\ssz HANNELS}\\
\hline \hline $\tilde\chi\tilde\chi$ & $f \bar f$ & $
s(h),s(H),s(A),s(Z)$
\\
&( $f$: Fermions)  & $t(\sff_{1,2}),u(\sff_{1,2})$
\\
& $H^\pm W^\mp$&  $ s(h),s(H),s(A),t(\chai),u(\chai)$
\\
&$hA,ZH$&  $ s(A),s(Z),t(\nti),u(\nti)$
\\ \hline
$\tilde\chi\tilde\tau_2$ & $\tau h,\tau H, \tau Z$ &
$s(\tau),t(\tilde\tau_{1,2}),u(\tilde\chi^0_i)$
\\
& $\tau A$ & $s(\tau),t(\tilde\tau_1),u(\tilde\chi^0_i)$
\\
& $\tau\gamma$ & $s(\tau),t(\tilde\tau_2)$
\\
& $\ntt H^-,\ntt W^-$ &
$s(\tau),t(\tilde\chi^\pm_i),u(\tilde\nu_{\tau})$
\\
\hline $\tilde\tau_2\tilde\tau_2$ & $\tau\tau$ &
$t(\tilde\chi^0_i),u(\tilde\chi^0_i)$
\\
\hline $\tilde\tau_2\tilde\tau_2^\ast$ & $hh,hH,HH,ZZ$ &
$s(h),s(H),t(\tilde\tau_{1,2}),u(\tilde\tau_{1,2}),{\rm PI}$
\\
& $AA$ & $s(h),s(H),t(\tilde\tau_1), u(\tilde\tau_1),{\rm PI}$
\\
&$H^+ H^-,W^+ W^-$ &
$s(h),s(H),s(\gamma),s(Z),t(\tilde\nu_{\tau}),{\rm PI}$
 \\
&$H^\pm W^\mp$ & $s(h),s(H),t(\tilde\nu_{\tau})$
\\
&$AZ$ & $s(h),s(H),t(\tilde\tau_1),u(\tilde\tau_1)$
\\
& $\gamma\gamma,\gamma Z$ & $t(\tilde\tau_2),u(\tilde\tau_2),{\rm
PI}$
\\
& $t\bar t,b\bar b$ & $s(h),s(H),s(\gamma),s(Z)$
\\
& $\tau\bar\tau$ & $s(h),s(H),s(\gamma),s(Z), t(\tilde\chi)$
\\
%& $u\bar u,d\bar d,e \bar e$ & $s(\gamma),s(Z)$
%\\
\hline
\end{tabular}
\end{center}
\end{table}

Bino-stau coannihilations come into play, when
$\Delta_{\tilde\tau_2}\lesssim0.25$. $\Omega_{\rm LSP}h^2$ is not
any more inverse proportional to $\sigv{\tilde\chi}{\tilde\chi}$
but to an effective $\sigv{}{}$ which includes, in addition,
$\sigv{\tilde\tau_2}{\tilde\chi[\tilde\tau_2]}$ with a weight
factor $\exp{\left(-[2]\Delta_{\tilde\tau_2}x^{-1}_{\rm
F}\right)}$, where $x^{-1}_{\rm F}=m_{\rm LSP}/T_{\rm F}$ $\sim25$
and $T_{\rm F}$ the freeze out temperature \cite{gelmini}.
Consequently, for given $m_{\rm LSP}$, $\Delta_{\tilde\tau_2}$
regulates the degeneracy amount. The strongest possible reduction
is achieved for $\Delta_{\tilde\tau_2}=0$.

The computation of $\Omega_{\rm LSP}h^2$ in this regime, can be
realized by using exclusively the low velocity expansion of
$\sigv{}{}$. The thermal average has been performed, following the
Ref. \cite{ellis2}. The included set of (co)annihilation processes
(in accord with the tables in Refs. \cite{roberto}) is shown in
the Table 3 (we apply the notation of Ref. \cite{cdm} and PI
stands for ``point interaction''). The relevant matrix elements
are evaluated with the help of the {\tt FeynCalc} package
\cite{feyncalc}. Our results are compared with {\tt micrOMEGAs},
in the Table 4 for 3 test points. Besides the values of
$\Omega_{\rm LSP}h^2$ and $x_{\rm F}^{-1}$, we display, also, the
contribution of all the channels, beyond $0.3\%$. As we can
observe, our differences are small in most cases.

The importance of CAM, in deriving an upper bound on $m_{\rm LSP}$
from Eq. (\ref{cdmb}), can be easily concluded from Fig. \ref{om}
(right plot), where we depict $\Omega_{\rm LSP}h^2$ versus $m_{\rm
LSP}$ for $\mu<0,\Delta_{\tilde\tau_2}=0$ and $m^{\rm
c}_b(M_Z)=2.68~{\rm GeV}$. $\Omega_{\rm LSP}h^2$ is calculated
with (thick solid line [crosses]) or without (faint solid line
[crosses]) the inclusion of bino-stau coannihilations using {\tt
micrOMEGAs} [our code]. We see that the reduction of $\Omega_{\rm
LSP}h^2$ caused by the CAM is dramatic (a factor of 10). This
increases the upper bound on $m_{\rm LSP}$ derived from Eq.
(\ref{cdmb}) from about $200~{\rm GeV}$ to $900~{\rm GeV}$.

\addtolength{\textheight}{1.cm}
\newpage

\begin{table}[!h]
\caption {\bf C\ssz OMPARISON OF \nsz {\boldmath $\Omega_{\rm
LSP}h^2$} C\ssz ALCULATIONS}
\begin{center}
\begin{tabular}{|l||c|c||c|c||c|c|}\hline
\multicolumn{7}{|c|}{\bf M\ssz ODEL \nsz P\ssz ARAMETERS}\\
\multicolumn{7}{|c|}{($\mu<0,~m^{\rm c}_b(M_Z)=2.68~{\rm GeV},
~A_0=0~{\rm GeV}$)}\\ \hline
$\tan\beta$&\multicolumn{2}{c||}{51.2}&\multicolumn{2}{|c||}{50.0}&
\multicolumn{2}{|c|}{49.4} \\
$M_{1/2}~({\rm GeV})$
&\multicolumn{2}{c||}{1877}&\multicolumn{2}{c||}{859}&
\multicolumn{2}{c|}{596}  \\
$m_0~({\rm GeV})$
&\multicolumn{2}{c||}{957}&\multicolumn{2}{c||}{463}&
\multicolumn{2}{c|}{357}  \\ \hline
$m_A~({\rm GeV})$&\multicolumn{2}{c||}{$466$}
&\multicolumn{2}{c||}{214}& \multicolumn{2}{c|}{130} \\
$m_{\rm LSP}~({\rm GeV})$ &\multicolumn{2}{c||}{865}
&\multicolumn{2}{c||}{386}& \multicolumn{2}{c|}{264} \\
$\Delta_{\tilde\tau_2}$&\multicolumn{2}{c||}{0.00}&\multicolumn{2}{c||}{0.05}&
\multicolumn{2}{c|}{0.10} \\ \hline \hline
\multicolumn{7}{|c|}{\bf \nsz {\boldmath $\Omega_{\rm LSP}h^2$}
C\ssz ALCULATIONS}\\ \hline
\multirow{2}{1in}{\bf C\ssz ODE \nsz :}&{\tt
micr-}&\multirow{2}{0.31in}{GLP}&{\tt
micr-}&\multirow{2}{0.31in}{GLP}&{\tt
micr-}&\multirow{2}{0.31in}{GLP} \\ &{\tt OMEGAs}& &{\tt OMEGAs}&
&{\tt OMEGAs}&\\ \hline \hline
\multicolumn{7}{|c|}{\bf P\ssz ROCESSES \nsz W\ssz HICH \nsz C\ssz
ONTRIBUTE \nsz M\ssz ORE \nsz T\ssz HAN \nsz {\boldmath
$0.3\%$}}\\ \hline \multicolumn{1}{|l||}{\bf P\ssz ROCESS} &
\multicolumn{6}{|c|}{\bf C\ssz ONTRIBUTION \nsz ({\boldmath
$\%$})}\\ \cline{2-7} \hline
$\tilde\chi\tilde\chi\rightarrow f\bar f$
&0.4&0.5&8.4&10.6&25.7&30.5\\ $\tilde\chi\tilde\chi\rightarrow
H^\pm W^\mp$ &0.8&0.7&9.4&10.1&18.6&19.6\\
$\tilde\chi\tilde\chi\rightarrow H Z$ &0.4&0.2&5.1&3.8&11.2&10.4\\
$\tilde\chi\tilde\chi\rightarrow h A$ &0.4&0.3&5.2&4.3&11.0&9.3\\
$\tilde\chi\tilde\chi\rightarrow h H$ &-&-&-&-&0.3&0.3\\
$\tilde\chi\tilde\chi\rightarrow A Z$ &-&-&-&-&0.3&0.2\\ \hline
$\tilde\chi\tilde\tau_2\rightarrow\tau
h$&1.4&0.4&8.0&6.9&7.1&6.1\\
$\tilde\chi\tilde\tau_2\rightarrow\tau
H$&3.0&2.8&4.8&4.6&2.6&2.4\\
$\tilde\chi\tilde\tau_2\rightarrow\tau
A$&3.0&3.3&4.7&5.3&2.7&2.8\\
$\tilde\chi\tilde\tau_2\rightarrow\nu_\tau
H^-$&5.8&6.0&8.5&9.1&4.4&4.4\\
$\tilde\chi\tilde\tau_2\rightarrow\nu_\tau W^- $&0.6&0.5&1.5
&1.3&0.9&0.7\\
$\tilde\chi\tilde\tau_2\rightarrow\tau\gamma$&7.9&7.7&11.9
&12.3&6.4&6.3\\ $\tilde\chi\tilde\tau_2\rightarrow\tau
Z$&2.7&2.7&4.8 &5.0&2.8&2.7\\ \hline
$\tilde\tau_2\tilde\tau_2\rightarrow\tau\tau$&19.0&19.2&5.5
&5.7&0.9&0.8\\ \hline
$\tilde\tau_2\tilde\tau_2^\ast\rightarrow hh$ &8.0&8.2&4.2
&3.9&0.8&0.6\\ $\tilde\tau_2\tilde\tau_2^\ast\rightarrow HH$
&3.0&3.2&0.9&0.9&-&-\\  $\tilde\tau_2\tilde\tau_2^\ast\rightarrow
AA$ &3.0&3.2&0.9&0.9&-&-\\
$\tilde\tau_2\tilde\tau_2^\ast\rightarrow
H^+H^-$&5.6&6.2&1.3&1.5&-&-\\
$\tilde\tau_2\tilde\tau_2^\ast\rightarrow H^\pm
W^\mp$&0.4&0.4&-&-&-&- \\
$\tilde\tau_2\tilde\tau_2^\ast\rightarrow
W^+W^-$&13.8&14.4&4.4&4.6&0.6&0.6\\
$\tilde\tau_2\tilde\tau_2^\ast\rightarrow ZZ$
&7.3&7.7&2.3&2.4&0.3&0.3\\
$\tilde\tau_2\tilde\tau_2^\ast\rightarrow\gamma\gamma$&5.7&5.3
&1.7&1.7&0.3&0.2\\ $\tilde\tau_2\tilde\tau_2^\ast\rightarrow\gamma
Z$&3.0&3.2&0.7&0.8&-&-\\ $\tilde\tau_2\tilde\tau_2^\ast\rightarrow
t\bar t$&4.1&3.0&4.5&3.5 &0.9&0.6
%\\ $\tilde\tau_2\tilde\tau_2^\ast\rightarrow \tau\bar\tau$& & & &&&
\\ \hline \hline
$x^{-1}_{\rm F}$ &25.97&26.03&25.51&25.48&24.82&24.84\\
$\Omega_{\rm LSP}h^2$ &0.221&0.205&0.221&0.221&0.220&0.216 \\
\hline
\end{tabular}
\end{center}
\end{table}

\addtolength{\textheight}{-1.cm}
\newpage

\section{P{\ftn HENOMENOLOGICAL} C{\ftn ONSTRAINTS}}
\label{pheno}

\hspace{.562cm} A lower bound on $m_{\rm LSP}$ can be derived by
imposing a number of phenomenological constraints \cite{nath,
baery, cmssm, baer}. These result from:

{\bf i.} The Higgs boson masses. The relevant for our analysis is
the $95\%$ c.l. LEP bound \cite{higgs} on the lightest CP-even
neutral Higgs boson, $h$ mass
\beq m_h\gtrsim114.4~{\rm GeV}\label{mhb} \eeq
which gives a [almost always the absolute] lower bound on $m_{\rm
LSP}$ for $\mu<[>]~0$. The SUSY contributions to $m_h$ are
calculated at two-loop by using the {\tt FeynHiggsFast} \cite{fh}
program included in {\tt micrOMEGAs} package \cite{micro}.

{\bf ii.} The deviation $\delta a_\mu$ of the muon anomalous
magnetic moment measured value, $a_\mu$, from its predicted in the
SM, $a^{\rm SM}_\mu$. The latter is not yet stabilized mainly due
to the instability of the hadronic vacuum polarization
contribution. According to the most updated \cite{davier}
evaluation of this contribution, the findings based on $e^+e^-$
data and on $\tau$-decay data are inconsistent with each other.
Combining these results with the recent experimental measurements
on $a_\mu$ \cite{muon}, we obtain the following $95\%$ c.l.
ranges:
\bea \mbox{\sf a)}\hspace{16.9pt}11.3\times10^{-10}\lesssim\delta
a_\mu&~~\mbox{and}~~&\mbox{\sf b)}~~\delta a_\mu\lesssim
56.1\times 10^{-10}~~~~\mbox{$e^+e^-$-based}\label{g2e}
\\\vspace*{19pt}
\mbox{\sf a)}~-11.6\times10^{-10}\lesssim\delta
a_\mu&~~\mbox{and}~~&\mbox{\sf b)}~~\delta a_\mu\lesssim30.4\times
10^{-10}~~~~\mbox{$\tau$-based}\label{g2t} \eea
The SUSY contribution to the $\delta a_\mu$ is calculated by using
the formulae of Ref. \cite{gmuon} in accord with {\tt micrOMEGAs}
\cite{micro} and the result is posi[nega]-tive for $\mu>[<]~0$. A
lower bound on $m_{\rm LSP}$ can be derived for $\mu>[<]~0$ from
Eq. (\ref{g2e}{\sf b}) [(\ref{g2t}{\sf a})] and an optimistic
upper bound for $\mu>0$ from Eq. (\ref{g2e}{\sf a}) which, however
is not imposed as an absolute constraint.
%, due to the former computational
%instabilities

{\bf iii.} The inclusive branching ratio of $b\rightarrow
s\gamma$, ${\rm BR}(b\rightarrow s\gamma)$. Taking into account
the recent experimental results \cite{cleo} on this ratio and
combining appropriately the experimental and theoretical involved
errors \cite{qcdm}, we obtain the following $95\%$ c.l. range:
\beq \mbox{\sf a)}~~1.9\times 10^{-4}\lesssim {\rm
BR}(b\rightarrow s\gamma)~~~~\mbox{and}~~~~
\mbox{\sf b)}~~{\rm BR}(b\rightarrow s\gamma)\lesssim 4.6 \times
10^{-4} \label{bsgb} \eeq
To calculate ${\rm BR}(b\rightarrow s\gamma)$ we used an updated
version of the code contained in the current version of {\tt
micrOMEGAs} \cite{bsgmicro}. This code represents a complete
update respect the one used in the first paper of Ref.
~\cite{qcdm}. The SM contribution is calculated using the Refs.
\cite{kagan}. The charged Higgs boson [SUSY] contribution is
evaluated by including the next-to-leading order QCD [SUSY
resummed] corrections and $\tan\beta$ enhanced contributions from
Refs. \cite{nlo}. A lower bound on $m_{\rm LSP}$ can be derived
for $\mu>[<]~0$ from Eq. (\ref{bsgb}{\sf a}) [(\ref{bsgb}{\sf b})]
with the latter being much more restrictive.

\section{C{\ftn ONCLUSIONS}-O{\ftn PEN} I{\ftn SSUES}}
\label{cncl}

\hspace{.562cm} Allowing $m^{\rm c}_b(M_Z)$ to vary in its 95$\%$
c.l. range, Eq. (\ref{mbrg}) and using the previous
Cosmo-Phenomenological restrictions, we can delineate the
parameter space of the model on the $m_{\rm
LSP}-\Delta_{\tilde\tau_2}$ plane in Figs. 4 and 5. For
simplicity, we do not show bounds from less restrictive
constraints, Eq. (\ref{g2e}{\tt b}) and (\ref{bsgb}{\tt a})
[(\ref{mhb})] for $\mu>[<]~0$.

\begin{figure}[t]
\hspace*{-.71in}
\begin{minipage}{8in}
\epsfig{file=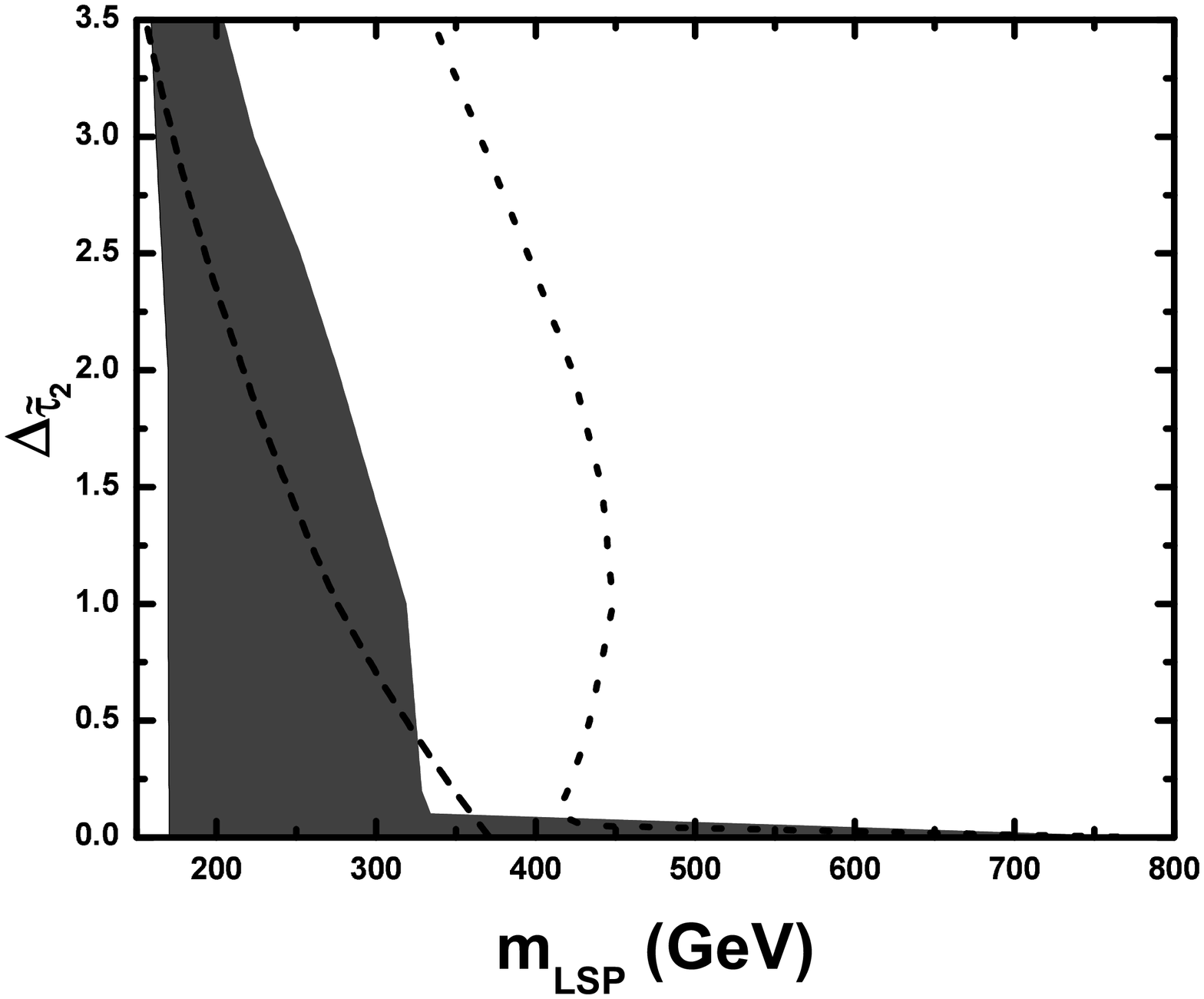,height=3.8in,angle=-90}
\hspace*{-1.37 cm}
\epsfig{file=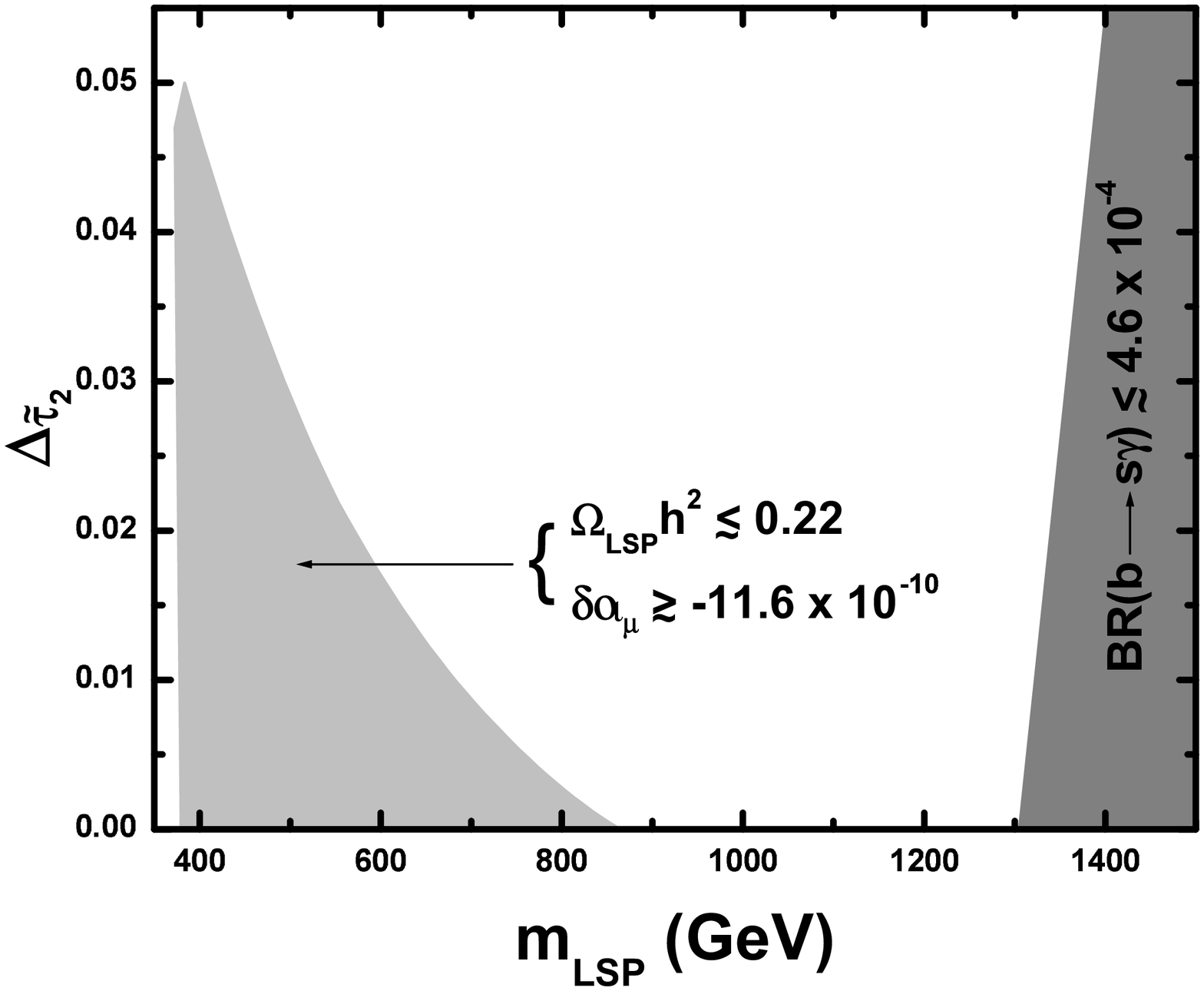,height=3.8in,angle=-90}
\hfill
\end{minipage}\vspace*{-.20in}
\begin{flushleft}
\begin{tabular}[!h]{ll}
%\hspace*{-.15in}
\hspace*{-.14in}
\begin{minipage}[t]{7.37cm}\caption[]{\sl The allowed area on the
$m_{\rm LSP}$-$\Delta_{\tilde\tau_2}$ plane for $\mu>0$ and
$2.68~{\rm GeV}\lesssim m^{\rm c}_b(M_Z)\lesssim 3.09~{\rm GeV}$.
Shown is, also, the [optimistic] upper bound from Eq.
[(\ref{g2e}{\sf a})] (\ref{cdmb}) with $2.61~{\rm GeV}\lesssim
m_b(M_Z) \lesssim3.15~{\rm GeV}$, [dashed] dotted
line.}\label{mup}
\end{minipage} &\begin{minipage}[t]{7.37cm}\caption[]{\sl Disconnected
areas on the $m_{\rm LSP}$-$\Delta_{\tilde\tau_2}$ plane for
$\mu<0$ and $2.68~{\rm GeV}\lesssim m^{\rm c}_b(M_Z)
\lesssim3.09~{\rm GeV}$ allowed by various restrictions indicated
in the plot.}\label{mun}
\end{minipage}
\end{tabular}
\end{flushleft}\vspace*{-.11in}
\end{figure}

For $\mu>0$, the allowed space of parameters (depicted in Fig.
\ref{mup}) is wide as regards both $\Delta_{\tilde\tau_2}$
(because of the $A$PE) and $m_{\rm LSP}$ (because of the CAM)
range. Especially,
\bea \Big(\mbox{From Eq. (\ref{mhb})}:\Big)\>\>\> 159~{\rm GeV}
\lesssim m_{\rm LSP}\lesssim720~[371]\>{\rm
GeV}\>\>\>\Big(:\mbox{From Eq. (\ref{cdmb}) [(\ref{g2e}{\sf
a})]}\Big) \nonumber \eea
(taking the less restrictive values of the various constraints).
The changes on the upper limit of the allowed area from a possible
inclusion of the SUSY corrections (sec. \ref{pole}) in the
$\Omega_{\rm LSP}h^2$ calculation are almost invisible. On the
contrary, let $\alpha_s(M_Z)$ vary in its $95\%$ c.l. experimental
range $0.1145-0.1225$ with the corresponding range of $m^{\rm
c}_b(M_Z)$ to be ($2.62-3.16)~{\rm GeV}$, the allowed area can
further enlarge until the dotted line in the Fig. 4.

On the other hand, for $\mu<0$, the competition of the various
Cosmo-Phenomenological constraints is more dangerous and, finally,
disastrous for this case, since:
\bea  \Big(\mbox{From Eq. (\ref{g2t}{\sf a})}:\Big)\>\>\>387~{\rm
GeV}\lesssim & m_{\rm LSP}\lesssim873\>\>\>{\rm GeV}\>\>\>&\Big(:
\mbox{From Eq. (\ref{cdmb})}\Big)\nonumber \\ \mbox{and}\>\>\> &
m_{\rm LSP}\gtrsim1305\>{\rm GeV} \>\>\>&\Big(: \mbox{From Eq.
(\ref{bsgb}{\sf b})}\Big)\nonumber\eea
Despite the strong presence of the CAM
($\Delta_{\tilde\tau_2}\lesssim0.05$) we are, evidently, left
without simultaneously allowed region, as it is shown in Fig.
\ref{mun}.

%%%%%%%%%%%%%%%%%%%%%%%%%%%%%%%%%%%%%%%%%%%%%%%%%%%%%%%%%%%%%%%%%%%%
%%%%%%%%%%%%%%%%%%%%%%%%%%%%%%%%%%%%%%%%%

Constructions similar to this presented in sec. \ref{model} may be
useful for other SUSY GUTs, too. The $SO(10)$ or $SU(5)$ SUSY GUTs
in their simple realization lead to complete \cite{raby, baery} or
$b-\tau$ \cite{nath, kazakov} YU, with the latter being viable
only for $\mu<0$ and consequently, quite disfavored from the
bounds of Eq. (\ref{g2e}). Also, the large values of $\tan\beta$
predicted in these models, can enhance the neutralino detection
rates with universal \cite{babu} or non universal \cite{nath}
asymptotic gaugino masses. Furthermore, the extra higgs fields
used in sec. \ref{model} have interesting consequences to the
inflation mechanism, as it is pointed out in Ref. \cite{jean2}.

%\newpage

\acknowledgments

\hspace{.562cm} We would like to thank Prof. G. Lazarides, for
close and instructive collaborations, from which this work 
is culled. We are, also,  grateful to {\tt micrOMEGAs} 
team, G. B\'{e}langer, F. Boudjema, A. Pukhov and 
A. Semenov, for providing us their 
${\rm BR}(b\rightarrow s\gamma)$ updated code and for
their patient correspondence to our attempt for a new 
comparison. C.P. wishes to thank Prof. S. Sarkar, for his
invitation to give this talk, G. Senjanovi\'c and P. Ullio for
useful discussions. C.P. was supported by EU under the RTN
contract HPRN-CT-2000-00152. M.E.G. acknowledges support 
from the `Funda\c c\~ao
para a Ci\^encia e Tecnologia' under contract SFRH/BPD/5711/2001
and project CFIF-Plurianual (2/91).

\def\ijmp#1#2#3{{\sl Int. Jour. Mod. Phys. }{\bf #1},~#3~(#2)}
\def\plb#1#2#3{{\sl Phys. Lett. }{B \bf #1},~#3~(#2)}
\def\zpc#1#2#3{{\sl Z. Phys. }{C \bf #1},~#3~(#2)}
\def\prl#1#2#3{{\sl Phys. Rev. Lett. }{\bf #1},~#3~(#2)}
\def\rmp#1#2#3{{\sl Rev. Mod. Phys. }{\bf #1},~#3~(#2)}
\def\prep#1#2#3{{\sl Phys. Rep. }{\bf #1},~#3~(#2)}
\def\prd#1#2#3{{\sl Phys. Rev. }{D \bf #1},~#3~(#2)}
\def\npb#1#2#3{{\sl Nucl. Phys. }{\bf B#1},~#3~(#2)}
\def\npps#1#2#3{{\sl Nucl. Phys. }{B (Proc. Sup.) \bf #1},~#3~(#2)}
\def\mpl#1#2#3{{\sl Mod. Phys. Lett.}{\bf #1},~#3~(#2)}
\def\ibid#1#2#3{{\sl ibid. }{\bf #1},~#3~(#2)}
\def\cpc#1#2#3{{\sl Comput. Phys. Commun. }{\bf #1},~#3~(#2)}
\def\astp#1#2#3{{\sl Astropart. Phys. }{\bf #1},~#3~(#2)}
\def\epjc#1#2#3{{\sl Eur. Phys. J. }{C \bf #1},~#3~(#2)}
\def\jhep#1#2#3{{\sl JHEP }{\bf #1},~#3~(#2)}

\end{document}